\documentclass[14pt]{article}
\usepackage{epsf}
\usepackage{color}
\usepackage{amsmath,amsfonts,amsbsy,amssymb}
\usepackage{indentfirst}
\usepackage{mathtools}
\usepackage{arydshln} 
\usepackage{framed} 

\newcommand{\fsl}[1]{\ensuremath{\mathrlap{\not{\phantom{#1}}}#1}}

\newcommand{\nn}{\nonumber}

\def\be{\begin{equation}}
\def\ee{\end{equation}}
\def\bse{\begin{subequations}}
\def\ese{\end{subequations}}
\def\bal{\begin{align}}
\def\ealn{\end{align}}
\def\tr{\text{tr}}
\def\bs{\boldsymbol}

\begin{document}

\begin{titlepage}

\def\slash#1{{\rlap{$#1$} \thinspace/}}

\begin{flushright} 

\end{flushright} 

\vspace{0.1cm}

\begin{Large}
\begin{center}

{\bf     
$SO(5)$ Landau Model  and 4D Quantum Hall Effect \\
in\\
 The $SO(4)$ Monopole Background 
}
\end{center}
\end{Large}

\vspace{1cm}

\begin{center}
{\bf Kazuki Hasebe}   \\ 
\vspace{0.5cm} 
\it{
National Institute of Technology, Sendai College,  
Ayashi, Sendai, 989-3128, Japan} \\ 

\vspace{0.8cm} 

{\today} 

\end{center}

\vspace{1.5cm}

\begin{abstract}
\noindent

\baselineskip=18pt


We investigate the $SO(5)$ Landau problem  in the $SO(4)$ monopole gauge field background by applying the techniques of the non-linear realization of quantum field theory.    
The $SO(4)$ monopole carries two topological invariants, the second Chern number and a generalized Euler number, specified by  the $SU(2)$  monopole and anti-monopole indices, $I_+$ and $I_-$. 
The  energy levels of the $SO(5)$ Landau problem are grouped into  $\text{Min}(I_+, I_-) +1$ sectors, each of which holds  Landau levels. 
In the $n$-sector,  
 $N$th  Landau level eigenstates constitute the $SO(5)$ irreducible representation with   $(p,q)_5=(N+I_+ + I_--n, N+n)_5$ whose function form is obtained  from  the $SO(5)$ non-linear realization matrix.  
In the $n=0$ sector, the emergent quantum geometry of the lowest Landau level is  identified as the fuzzy four-sphere with radius being proportional to the  difference between $I_+$ and $I_-$.  
The Laughlin-like wavefunction is constructed  by imposing the $SO(5)$ lowest Landau level projection to the many-body wavefunction made of the Slater determinant.  
 We also analyze the relativistic version of the $SO(5)$ Landau model to demonstrate the Atiyah-Singer index theorem in the  $SO(4)$ gauge field configuration.

\end{abstract}

\end{titlepage}

\newpage 

\tableofcontents

\newpage 

\section{Introduction}

 More than forty years ago, Yang introduced the $SU(2)$ monopole that epitomizes beautiful  topological features of  non-Abelian gauge field   \cite{Yang-1978-1, Yang-1978-2}. The $SU(2)$ monopole  on $S^4$  realizes  a natural non-Abelian generalization of the $U(1)$ principal fibre of the Dirac monopole on $S^2$ \cite{Dirac-1931}, and the $SU(2)$ monopole charge exemplifies a physical manifestation of the second Chern number. Not only for its elegant mathematical structure, the $SU(2)$ monopole  found its physical applications in the $SO(5)$ Landau model and 4D quantum Hall effect \cite{Zhang-Hu-2001}, which, from a modern point of view, is the first theoretical model of a  topological insulator  in  higher dimension. The underlying geometry of the system is the nested quantum Nambu geometry that does not have any counterpart in  classical geometry \cite{Hasebe-2020-1}, which renders  the system to be  quite unique also in  view of  the non-commutative geometry \cite{Hasebe-2014-1,  Hasebe-2017}. Tensor-type Chern-Simons theories are proposed as  effective field theories \cite{Hasebe-2014-1, Hasebe-2017} that naturally induce a generalized fractional statistics of extended objects \cite{WuZee1988, TzeNam1989, Hasebe-2020-3}.  
 The theoretical formulation of the quantum Hall effect has now been generalized to even higher dimensions \cite{Karabali-Nair-2002, Bernevig-Hu-Toumbas-Zhang-2003, Hasebe-Kimura-2003, Nair-Daemi-2004, 
Jellal-2005,  Hasebe-2010-2,  Balli-Behtash-Kurkcuoglu-Unal-2014, Hasebe-2014-2, Karabali-Nair-2016,  Coskun-Kurkcuoglu-Toga-2017,  Heckman-Tizzano-2018} and supersymmetric versions \cite{Hasebe-2004, Hasebe-2008}.

In recent years, studies of the higher D topological phases  took a new turn.   
The idea of the synthetic dimension and artificial gauge field allowed researchers to access higher dimensional  topological phases  with tabletop  experiments. 
 The artificial $SU(2)$ monopole gauge field has been  implemented in systems such as cold atoms   \cite{Sugawa-et-al-2018} and  meta-materials  \cite{Ma-Bi-et-al-2021}. 
For topological features specific to the 4D quantum Hall effect, a number of experiments have been proposed  in  cold atoms \cite{Price-et-al-2015,  Price-et-al-2016}, photonics \cite{Ozawa-Price-et-al-2016}, circuit  \cite{Wang-Price-Zhang-Chong-2020} and acoustics \cite{Chen-Zhu-Tan-Wang-Ma-2021}, and  several theoretical predictions have already been confirmed \cite{Lohse-et-al-2018,  Zilberberg-et-al-2018}.  Along with the developments, a 5D  Weyl semi-metal  with an  $SU(2)$ monopole and $SU(2)$ anti-monopole  structure in the momentum space  has  been proposed  \cite{Lian-Zhang-2016,Lian-Zhang-2017} and reported  to  host higher order topological insulators \cite{Hashimoto-Kimura-Wu-2017, Hashimoto-Matsuo-2020}.  
Partially inspired by the recent progress of higher D topological physics, we present a formulation of the 4D quantum Hall effect with an $SO(4)$ gauge structure. The $SO(4)\simeq SU(2)\otimes SU(2)$ group is only the semi-simple group  among all of the $SO(n)$ groups, and  
 the $SO(4)$ monopole can be regarded as a ``composite''  of  the $SU(2)$ monopole and the $SU(2)$ anti-monopole.   This notable structure is  significant in  the perspective of the topological insulator, because with  $SU(2)$ monopole and $SU(2)$ anti-monopole in the same magnitude,  the system may realize a non-chiral topological phase in a higher dimension.
 This feature is quite analogous to that  of the quantum spin Hall effect  \cite{Kane-Mele-2005, Bernevig-Zhang-2005, Sheng-Weng-Sheng-Haldane-2006}.\footnote{For a time-reversal symmetric 3D topological insulator with Landau levels, one may consult Refs.\cite{Li-Wu-2013, Li-Zhang-Wu-2013}.}   
Also in  perspectives of  the string theory, the non-chiral topological insulator is  interesting. 
  The formerly constructed even D quantum Hall systems  are all chiral that   correspond to the chiral superstring theory known as type II, while   
 non-chiral quantum Hall systems realize  rare set-ups that  correspond to the non-chiral  superstring theory known as type I \cite{Ryu-Takayanagi-2010-1, Ryu-Takayanagi-2010-2}. 

The $SO(4)$ gauge structure naturally appears in the context of  the 4D quantum Hall effect.\footnote{Strictly speaking,  the universal cover of  $SO(4)$, $i.e.$, $Spin(4)$, is adopted as the gauge group.}     
 In the set-up of the Landau models,  the gauge group  is adopted to be equal to the holonomy group of the base-manifold (see \cite{Hasebe-2010, Karabali-Nair-review} as  reviews). 
For the $SO(5)$ Landau model, the basemanifold is $S^4$ whose holonomy group is $SO(4)$,  and   in former researches,  
 one $SU(2)$  of the $SO(4)\simeq SU(2)\otimes SU(2)$ was adopted as the  gauge group.
Notably, Yang applied the method of separation of variables in solving the  differential equation of the $SO(5)$ Landau problem in the $SU(2)$ monopole background  and successfully  derived the eigenvalues and the eigenfunctions \cite{Yang-1978-2}.\footnote{Such monopole harmonics are known as the $SU(2)$ monopole harmonics, but  in the present paper, we   refer to the eigenstates as the $SO(5)$ monopole harmonics with emphasis on their $SO(5)$ covariance.} 
Though the analysis of the $SO(4)$ case is obviously significant,  it is still left unexplored. It may be because  the Landau problem in the $SO(4)$ monopole background is far more  complicated compared to the $SU(2)$ case. 
 To overcome such technical difficulties, we  adopt the techniques of non-linear realization.   
While the non-linear realization technique has been  developed in quantum field theory \cite{Coleman-Wess-Zumino-1969,CallanJr-Coleman-Wess-Zumino-1969,Salam-Strathdee-1982}, the non-linear realization is closely related to quantum mechanical systems with gauge symmetries \cite{Nair-book}  and has been successfully applied to  recent analyses of the Landau models \cite{Karabali-Nair-2002,Nair-Daemi-2004, Hasebe-2015, Hasebe-2018}.  
We use this method and completely solve the $SO(5)$ Landau model in the $SO(4)$ monopole background. With newly obtained monopole harmonics, we unveil particular properties of the $SO(5)$ Landau model and 4D quantum Hall effect. 

The paper is organized as follows. In Sec.\ref{sec:so3landaumodel}, we present a brief review about the non-linear realization of the $SO(3)$ Landau model.  Sec.\ref{sec:so5landaumodel} explains the Yang $SU(2)$ monopole  in a modern notation and  derives  a general form of the $SO(5)$ matrix generators.    
In Sec.\ref{sec:so5monoharnl},  we exploit the non-linear realization for the $SO(5)$ group.  
The $SO(5)$ Landau problem in the $SO(4)$ monopole background is investigated in Sec.\ref{sec:so5landauinso4}.   
In Sec.\ref{sec:noncommgeo}, we identify 
the non-commutative geometry and construct a Laughlin-like many-body wavefunction.  The relativistic  Landau model is discussed in Sec.\ref{sec:relativisticlandau} to demonstrate the Atiyah-Singer index theorem for the $SO(4)$ gauge field. Sec.\ref{sec:summary} is devoted to summary and discussions.

\section{$SO(3)$ monopole harmonics and non-linear realization}\label{sec:so3landaumodel}

\begin{figure}[tbph]
\hspace{-1.0cm}
\includegraphics*[width=180mm]{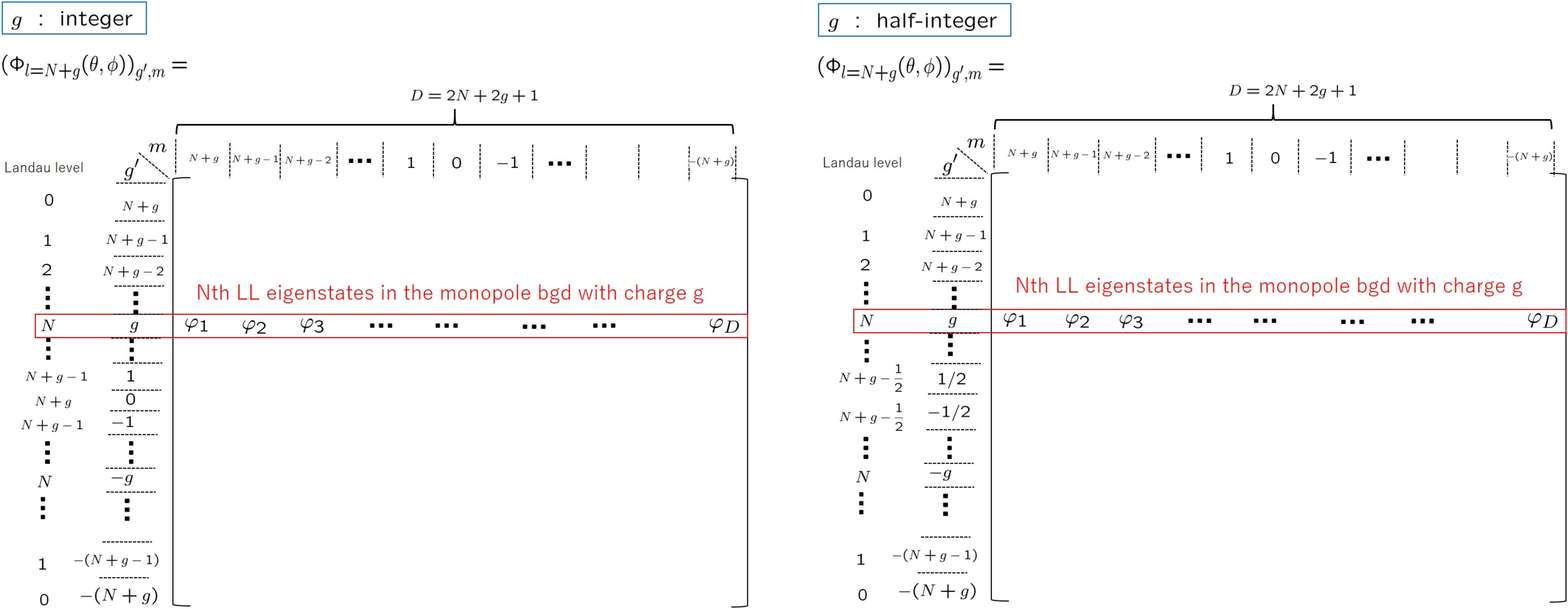}
\caption{ The $N$th Landau level eigenstates 
 are realized as the components of the red enclosed $(N+1)$th column of  the non-linear realization. }
\label{so3cosetfunc.fig}
\end{figure}

The monopole harmonics are known as the eigenstates of the $SU(2)$ Casimir of the  angular momentum in the Dirac monopole background.  In the Dirac gauge, the monopole gauge field is given by 
\be
A_i=-g\frac{1}{r(r+z)}\epsilon_{ij 3}x_j,  
\label{diracformalismgaugefields}
\ee
and the corresponding magnetic field is derived as 
\be
B_i=\epsilon_{ijk}\partial_j A_k=g\frac{1}{r^3}x_i. 
\ee
Here, $g$ takes an integer or a half-integer due to the Dirac quantization condition.\footnote{
The $U(1)$ monopole charge is given by 
\be
c_1 =\frac{1}{2\pi}\int_{S^2}B=2g, \label{intc1chern}
\ee
which represents the first Chern number of integer value. The result (\ref{intc1chern}) is consistent with the  fact that $g$ is either an integer or a half-integer.  
}
The covariant angular momentum operators are constructed as 
\be
\Lambda_i=-i\epsilon_{ijk}x_j(\partial_k+iA_k),  
\ee
and the total angular momentum operators are 
\be
L^{(g)}_i=\Lambda_i+r^2 B_i. \label{lisumlamb} 
\ee
In detail, (\ref{lisumlamb}) is given by 
\be
L^{(g)}_3=L^{(0)}_3+g,  ~~~~L^{(g)}_m=L^{(0)}_m+g\frac{1}{r+x_3}x_m~~(m=1,2), 
\ee
with
\be
L^{(0)}_i=-i\epsilon_{ijk}x_j\partial_k.
\ee
We introduce a non-linear realization of the $SU(2)$ group for the coset $SU(2)/U(1)$ as\footnote{When $l=1/2$,  Eq.(\ref{cosetrepsu2}) is represented as 
\be
\Phi_{1/2}(\theta, \phi) = e^{i\frac{\theta}{2}\sum_{m,n=1}^2 \epsilon_{mn}y_m(\phi)\sigma_n} =\begin{pmatrix}
\cos\frac{\theta}{2} & \sin\frac{\theta}{2}e^{-i\phi} \\
-\sin\frac{\theta}{2} e^{i\phi} & \cos\frac{\theta}{2}
\end{pmatrix} =\frac{1}{\sqrt{2(1+x_3)}} 
\begin{pmatrix}
1+x_3 & x_1-ix_2 \\
-x_1-ix_2 & 1+x_3 
\end{pmatrix} \label{nonline1/2}
\ee
where 
\be
x_1=\sin\theta \cos\phi, ~~x_2=\sin\theta\sin\phi,~~x_3=\cos\theta. 
\ee
One may readily check that $\Phi_{1/2}(\theta, \phi)$ (\ref{nonline1/2}) satisfies (\ref{covmatu1}):  
\be
L_i^{(g=\frac{1}{2}\sigma_3)}\Phi_{1/2}(\theta, \phi) =\Phi_{1/2}(\theta, \phi)~\frac{1}{2}\sigma_i. 
\ee
}  
\be 
\Phi_l(\theta, \phi) 
=e^{i\theta\sum_{m,n=1}^2 \epsilon_{mn}y_m(\phi)S_n^{(l)}}, \label{cosetrepsu2}
\ee 
where  
\be
y_1\equiv \cos\phi, ~~~y_2\equiv \sin\phi,
\ee
and $S_i^{(l)}$ denote the $SU(2)$ matrices of spin magnitude $l$ with their third component being     
\be
S_z^{(l)}=\text{diag}(l, l-1, l-2, \cdots, -l).
\ee
We see that   the non-linear realization (\ref{cosetrepsu2}) is a $(2l+1)\times (2l+1)$ matrix that satisfies 
\be
L_i^{(g=S_z^{(l)})} \Phi_l(\theta, \phi) =\Phi_l(\theta, \phi) ~S^{(l)}_i. \label{covmatu1}
\ee
By denoting the components of  $\Phi_l(\theta,\phi)$ as  
\be
\varphi_{l,m}^{(g)}(\theta, \phi) \equiv (\Phi_l(\theta, \phi))_{g,m},~~~~~(g,m=l,l-1,l-2,\cdots, -l+1, -l) \label{varphimono}
\ee
Eq.(\ref{covmatu1}) is recast into the following form 
\be
L_i^{(g)} \varphi_{l,m}^{(g)} =\sum_{m'=-l}^l \varphi_{l, m'}^{(g)} ~(S_i^{(l)})_{m' m}, 
\ee
and then 
\be
{L_i^{(g)}}^2 \varphi_{l,m}^{(g)}  =\sum_{m'=-l}^l \varphi_{l, m'}^{(g)} ~({S_i^{(l)}}^2)_{m' m}=l(l+1) \varphi_{l, m}^{(g)},  \label{so3caseige}
\ee
which indicates that $\varphi_{l,m}^{(g)}(\theta, \phi)$ realize the  monopole harmonics introduced in \cite{Wu-Yang-1976}. 
With normalization factors, the normalized monopole harmonics are expressed as    
\be
\sqrt{\frac{2l+1}{4\pi}}~\varphi_{l,m}^{(g)}(\theta,\phi). 
\label{so3normono} 
\ee
Notice that the non-linear realization (\ref{cosetrepsu2}) is factorized as 
\be 
\Phi_l(\theta, \phi) =e^{-i\phi S_z^{(l)}}e^{i\theta S_y^{(l)}}e^{i\phi S_z^{(l)}}
=D_l(\phi,-\theta,-\phi). \label{cosetrepsu2dfunc}
\ee 
Here, $D$ is  Wigner's $D$-functions (see \cite{Shnir-book} for instance): 
\be
D_l(\chi, \theta, \phi) =e^{-i\chi S_z^{(l)}}e^{-i\theta S_y^{(l)}}e^{-i\phi S_z^{(l)}}. 
\ee
Equation (\ref{varphimono}) is equal to $D_{l}(\phi, -\theta, -\phi)_{g,m}= d_{l,g,m}(-\theta)~e^{i(m-g)\phi}$ with $d_{j,m,m'}$ being  Wigner's small $D$-matrix:\footnote{The explicit form of (\ref{matrissywig}) is given by 
\be 
d_{j,m,m'}(\theta)
=(-1)^{m-m'}\sqrt{\frac{(j+m)!(j-m)!}{(j+m')!(j-m')!}}\biggl(\cos\frac{\theta}{2}\biggr)^{{m+m'}}\biggl(\sin\frac{\theta}{2}\biggr)^{{m-m'}} P_{j-m}^{(m-m',~ m+m')}(\cos\theta), 
\label{explicitsmalldfunc}
\ee
%
where $P^{(\alpha,\beta)}_n(x)$ stand for the Jacobi polynomials: 
\be
P_n^{(\alpha, \beta)}(x)=\frac{(-1)^n}{2^n n!} (1-x)^{-\alpha}(1+x)^{-\beta} \frac{d^n}{dx^n} (1-x)^{n+\alpha}(1+x)^{n+\beta}.  
\ee
}   
\begin{equation}
d_{j,m,m'}(\theta)=(e^{-i\theta S^{(j)}_y})_{m, m'}. \label{matrissywig}
\end{equation}
With  the monopole harmonics that satisfy (\ref{so3caseige}), it is now feasible to 
 solve  the $SO(3)$ Landau problem on a sphere \cite{Haldane-1983, Wu-Yang-1976}:  
\be
H=\frac{1}{2M}\sum_{i=1}^3 {\Lambda_i}^2 =\frac{1}{2M}(\sum_{i=1}^3 {L_i^{(g)}}^2 -r^4\sum_i {B_i}^2)=\frac{1}{2M}(\sum_{i=1}^3 {L_i^{(g)}}^2 -g^2). \label{so3landaumoham}
\ee
While  $l$ was assumed to be a given quantity,  
the input parameter in the Landau Hamiltonian is  the monopole charge $g$, and then  $l$ should be determined by $g$.   
In the following  
 we assume $g\ge 0$ for simplicity. The $SU(2)$ spin index  $l$ is greater than or equal to $g$, and so  $l$ starts from $g$ (not from $0$). Therefore,   the Landau level index $N$  may be identified as  
\be 
N\equiv l-g=0,1,2,\cdots. \label{nl-g}
\ee 
We then identify the $SU(2)$ spin index $l$ of the non-linear realization (\ref{cosetrepsu2})  as  
\be
l =N+g.  \label{n+gl}
\ee
From (\ref{varphimono}), we can now derive  
 the $(N+1)$th column of $\Phi_{l=N+g}(\theta, \phi)$ as the set of the $N$th Landau level eigenstates: 
\be
\Phi_{l=N+g, g, m}=\varphi_{l=N+g,m}^{(g)}~~~(m=l,l-1,l-2,\cdots, -l). \label{eigenstatesfrompsi}
\ee
See Fig.\ref{so3cosetfunc.fig}. 
Equation (\ref{so3caseige}) implies that  the eigenenergy of  (\ref{so3landaumoham}) is given by   
\be
E_N =\frac{1}{2M}({S_i^{(l=N+g)}}^2 -g^2) =\frac{1}{2M}(l(l+1)|_{l=N+g} -g^2) =\frac{1}{2M}(N(N+1)+g(2N+1)),  \label{landauso3}
\ee
and  
 (\ref{eigenstatesfrompsi}) denotes the $N$th Landau level  eigenstates. 
Notice that we first identified the Landau level eigenstates as the non-linear realization, and later we derived the Landau energy levels from the $SU(2)$ covariance of the non-linear realization.    

Let us summarize the essence of the non-linear realization technique. 
 Once the non-linear realization was constructed, we can read off the lowest and higher Landau level eigenstates from its matrix elements.  
In the construction of the non-linear realization (\ref{cosetrepsu2}), what we needed was just the higher spin matrices. 
The explicit form of the  higher spin matrices has been known, but even if we did not know them, we can  derive them   by sandwiching the angular momentum operators with some appropriate irreducible representation, say,     the lowest Landau level (LLL) eigenstates.\footnote{
Using the LLL eigenstates $\varphi_{g, m}^{(g)}$, we can  construct the higher spin matrices with spin magnitude $g$ by the formula:   
\be
\frac{2g+1}{4\pi}\int_{S^2} d\Omega_2 ~{\varphi_{g,m}^{(g)}}^*L_i^{(g)} \varphi_{g,m'}^{(g)} =(S_i^{(g)})_{mm'}. 
\ee
}  
In the following sections, we apply these observations for solving  the $SO(5)$ Landau problem in the $SO(4)$ monopole background. 

\section{$SO(5)$ matrix generators from Yang's monopole harmonics}\label{sec:so5landaumodel}
 
 We first need  to derive the matrix generators of  arbitrary $SO(5)$ irreducible representations. Fortunately, Yang already derived  a complete basis set of the $SO(5)$ irreducible representations as the $SO(5)$ monopole harmonics \cite{Yang-1978-2}. Sandwiching the $SO(5)$ angular  momentum operators with the $SO(5)$ monopole harmonics,   we can in principle derive the $SO(5)$ matrix generators of arbitrary representations. 
In this section,  we review Yang's work  with a modern notation \cite{Hasebe-2020-1} and derive  a general matrix form of the $SO(5)$  generators.

\subsection{Basics of the $SO(5)$ representation}

\begin{figure}[tbph]
\center
\includegraphics*[width=90mm]{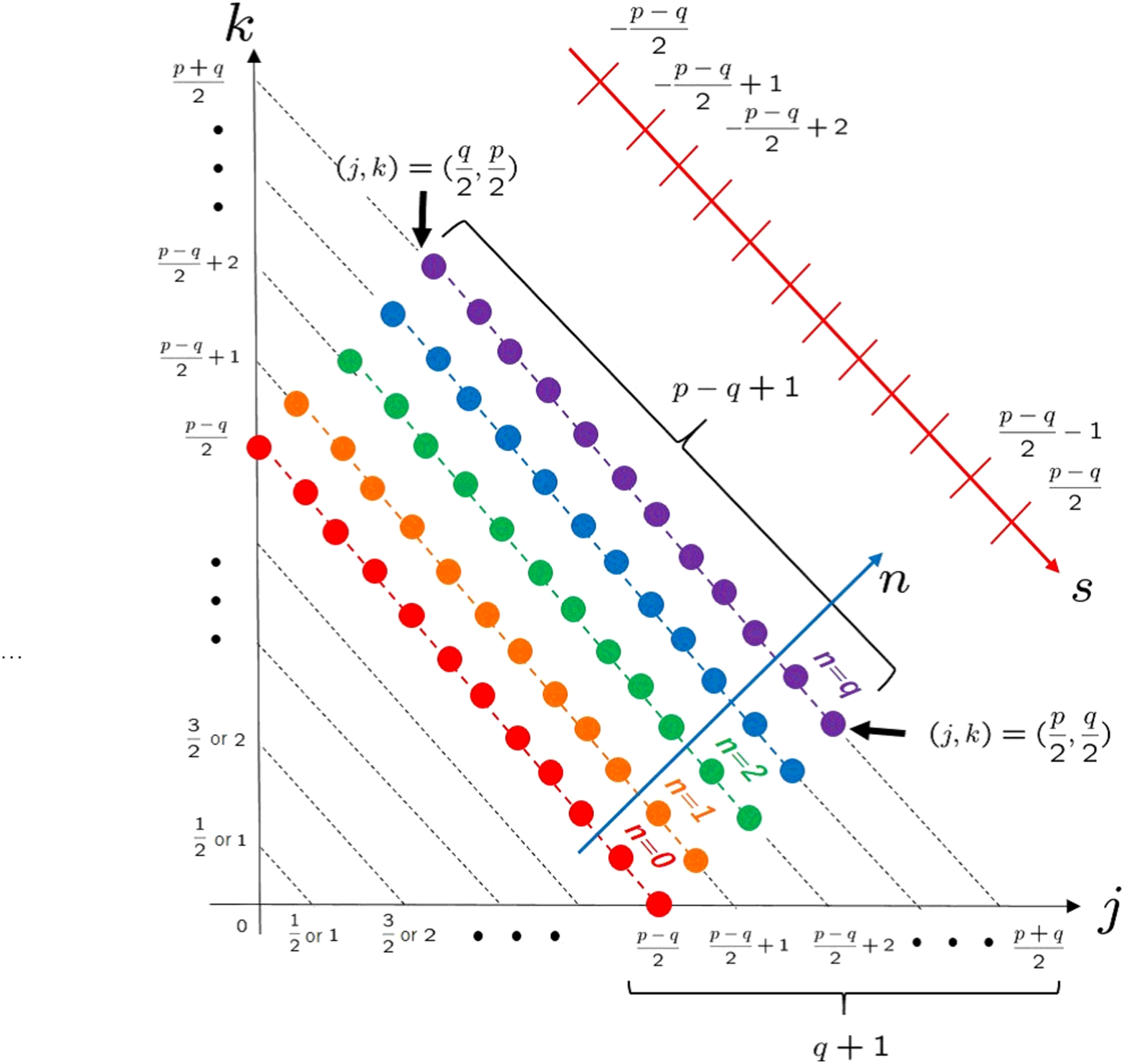}
\caption{ Each of the filled circles represents an $SO(4)$ irreducible representation.   The $SO(4)$ irreducible representations represented by the filled circles amount to the $SO(5)$ irreducible representation $(p,q)_5$. (Taken from \cite{Hasebe-2020-1}.)   }
\label{so5diagram.fig}
\end{figure}

The $SO(5)$ algebra holds   two non-negative integer Casimir indices, $p$ and $q$ $(p\ge q)$;  
the $SO(5)$ Casimir eigenvalue for the  $SO(5)$ irreducible representation, $(p,q)_5$,  is  given by  
\be
\lambda(p, q)=\frac{1}{2}p^2+\frac{1}{2}q^2 +2p+q, 
\label{lambso5gene}
\ee
and the corresponding dimension is 
\be
D(p, q)=\frac{1}{6}(p+2)(q+1)(p+q+3)(p-q+1). 
\label{so5degelandau}
\ee
The $SO(4) \simeq SU(2)\otimes SU(2)$ subgroup decomposition 
is given by [Fig.\ref{so5diagram.fig}] 
\be
(p, q)_5 =\bigoplus_{0\le n \le q} ~\bigoplus_{-\frac{p-q}{2} \le s \le {\frac{p-q}{2}}} ~
(j,k)_4, 
\ee
where  
\be
(j,k)_4 ~\equiv~ (\frac{n}{2}+\frac{p-q}{4}+\frac{s}{2}, ~\frac{n}{2}+\frac{p-q}{4}-\frac{s}{2})_4. 
\label{jkind}
\ee
The symbols,  $j$ and $k$, denote the bi-spin indices of the $SO(4)\simeq SU(2)\otimes SU(2)$ group, while $n=j+k-\frac{p-q}{2} ~(= 0, 1, 2, \cdots, q)$ and $s=j-k~(=-\frac{p-q}{2}, -\frac{p-q}+1, -\frac{p-q}{2}+2, \cdots ,\frac{p-q}{2})$ indicate the Landau level index and the chirality parameter in  the  $SO(4)$ Landau model \cite{Hasebe-2020-1}.   The  notations, $(j, k)_4$ and $[n, s]$, are both useful according to context and we  hereafter utilize them 
interchangeably:  
\be
(j,k)_4~~\longleftrightarrow~~[n,s]. \label{jknsinter}
\ee
Let us call 
 the oblique lines in Fig.\ref{so5diagram.fig} specified by  $j+k=n+\frac{p-q}{2}$  the $SO(4)$ lines.  Each filled circle represents an $SO(4)$ irreducible representation $(j, k)_4$ with dimension $(2j+1)(2k+1)$. On the $n$th $SO(4)$ line, there are $(p-q+1)$ $SO(4)$ irreducible representations and   the total dimension of those $SO(4)$ irreducible representations is counted as  
\be
d(n, p-q) \equiv \sum_{ -\frac{p-q}{2} \le s \le {\frac{p-q}{2}} } (2j+1)(2k+1) =\frac{1}{6}(p-q+1)((p-q)^2+(6n+5)(p-q)+6(n+1)^2).
\ee
As depicted in Fig.\ref{so5diagram.fig}, the $SO(4)$ irreducible representations on the $(q+1)$ $SO(4)$ lines $(n=0,1,2,\cdots, q)$ constitute the $SO(5)$ irreducible representation $(p, q)_5$: 
\be
\sum_{n=0}^q d(n, p-q)=D(p,q), 
\ee
where $D(p,q)$ is given by  (\ref{so5degelandau}). 

\subsection{$SO(5)$ monopole harmonics in the $SU(2)$ background}\label{subsec:so5ori}

In the Dirac gauge,  the $SU(2)$ anti-monopole gauge field \cite{Zhang-Hu-2001} is represented as 
\be
A_m=-\frac{1}{r(r+x_5)}\bar{\eta}_{mn}^i x_n S_i~~(m ,n=1, 2, 3, 4) ,~~~~A_5=0, \label{zhanghusu2gauge}
\ee
where  
$S_i$ $(i=1,2,3)$ denote the $SU(2)$ matrix of the spin $I/2$ representation, 
\be
S_iS_i =\frac{I}{2}(\frac{I}{2}+1)\bs{1}_{I+1}, 
\ee
and $\bar{\eta}_{mn}^i$ signifies  the  't Hooft symbol: 
\be
\eta_{mn}^i\equiv \epsilon_{mn i4} +\delta_{m i}\delta_{n 4}-\delta_{m 4}\delta_{n i}, ~~\bar{\eta}_{mn}^i\equiv \epsilon_{mn i4} -\delta_{m i}\delta_{n 4}+\delta_{m 4}\delta_{n i}. \label{defthooftsymbols}
\ee
We construct the covariant angular momentum operators as   
\be
\Lambda_{ab}=-ix_aD_b+ix_bD_a, ~~~~~(D_a=\partial_a +iA_a)
\ee
and the total $SO(5)$ angular momentum operators as  
\be
L_{ab}=
\Lambda_{ab}+
r^2F_{ab}.  \label{angso5tot}
\ee
The field strength, $F_{ab}=\partial_a A_b-\partial_b A_a +i[A_a, A_b]$ $(a,b=1,2,3,4,5)$, is derived as\footnote{The non-trivial topology of the $SU(2)$ monopole field configuration  is accounted for by  
\be
\pi_3(SU(2)) \simeq \mathbb{Z},  
\ee
and the corresponding second Chern number 
is evaluated as 
\be
c_2 =  \frac{1}{8\pi^2}\int_{S^4}\tr ~F^2=-\frac{1}{6}I(I+1)(I+2), \label{chernnumberantimo}
\ee
where $F=\frac{1}{2}F_{ab}dx_a\wedge dx_b$ with (\ref{constfieldstreng}). }  
\be
F_{mn}=-\frac{1}{r^2}x_m A_n+\frac{1}{r^2}x_n A_m+\frac{1}{r^2}\bar{\eta}^i_{mn}S_i,~~F_{m 5}=-F_{5m}=\frac{1}{r^2}(r+x_5)A_m, 
\label{constfieldstreng}
\ee
and (\ref{angso5tot}) is given by 
\be
L_{mn}={{L}}^{(0)}_{mn} +\bar{\eta}_{mn}^i S_i, ~~~L_{m 5} 
=L_{m 5}^{(0)}-\frac{1}{r+x_5}\bar{\eta}^i_{mn}x_n S_i, 
\label{detailantimonoso5gene}
\ee
where $L^{(0)}_{ab}$ denote the  $SO(5)$ free angular momentum operators: 
\be
L^{(0)}_{ab}=-ix_a\partial_b+ix_b\partial_a. 
\ee
Now the eigenvalue problem of the $SO(5)$ Casimir operator  reads 
\be
\sum_{a<b=1}^5 {L_{ab}}^2 ~\bs{\psi} =\lambda ~\bs{\psi}. 
\label{so5landaueigenval}
\ee
Yang showed that  with a given $SU(2)$ monopole index $I$, $p$ and $q$ are related as 
\be
p-q=I , 
\ee
or  
\be
(p, q)_5=(N+I, N)_5. \label{idenpanni}
\ee
Here $N$ denotes a non-negative integer value that corresponds to the Landau level of  the $SO(5)$ Landau model \cite{Zhang-Hu-2001}.      
Substituting (\ref{idenpanni}) into (\ref{lambso5gene}) and (\ref{so5degelandau})  respectively, we readily obtain   the $SO(5)$ Casimir eigenvalues of (\ref{so5landaueigenval}) 
and the   degeneracies as 
\begin{subequations}
\begin{align}
&\lambda(N+I, N)=N^2+N(I+3)+\frac{1}{2}I(I+4),   \\
&D(N+I, N)=\frac{1}{6}(N+1)(I+1)(I+N+2)(I+2N+3). 
\label{so5degelandaunth}
\end{align}
\end{subequations}
Thus, once the identification (\ref{idenpanni}) was established, the derivation of the eigenvalues is  an easy task, but the derivation of the eigenstates is another story.  Yang used the method of the separation of variables  for solving the differential equation  (\ref{so5landaueigenval}) \cite{Yang-1978-2}.  We will not  here repeat that derivation but just write down the results in a modern notation \cite{Hasebe-2020-1}.  
With the polar coordinates on a four-sphere (with unit radius)     
\begin{align}
&x_1=\sin\xi\sin\chi\sin\theta\cos\phi,~~x_2=\sin\xi\sin\chi\sin\theta\sin\phi,~~x_3=\sin\xi\sin\chi\cos\theta,\nn\\
&x_4=\sin\xi\cos\chi,~~x_5=\cos\xi, \nn \\ 
&~~~~~~~~(0\le \xi \le \pi,  ~~~0\le \chi \le \pi, ~~~0\le \theta \le \pi, ~~~0\le \phi < 2\pi), \label{polarx1tox5}
\end{align}
the normalized $SO(5)$ monopole harmonics are represented as\footnote{
 The orthonormal relation for the $SO(5)$ monopole harmonics is given by 
\be
\int d\Omega_4 ~\bs{\psi}_{N; j, m_j ;k, m_k}(\Omega_4)^{\dagger}~\bs{\psi}_{N'; j', m'_j ;k', m'_k}(\Omega_4) 
=\delta_{NN'}\delta_{jj'}\delta_{kk'}\delta_{m_j m_j'}\delta_{m_k m'_k},
\ee
where 
\be
d\Omega_4 =\sin ^3\xi~\sin ^2\chi~\sin\theta ~d\xi d\chi d\theta d\phi. 
\ee
}   
\be
\bs{\psi}_{N; j, m_j ;k, m_k}(\Omega_4) 
= G_{N,j,k}(\xi)\cdot \bs{Y}_{j, m_j; k, m_k}(\Omega_3), ~~~~(\Omega_3=(\chi,\theta,\phi))
 \label{x5so5monopolehamonicsnorm}
\ee
where 
\begin{subequations}
\begin{align}
&G_{N,j,k}(\xi) = (-1)^{2j+1}~\sqrt{N+\frac{I}{2}+\frac{3}{2}}~ \frac{1}{\sin\xi}~ d_{N+\frac{I}{2}+1, -j+k, j+k+1} (\xi), 
\label{exdefgmfunc} \\
&\bs{Y}_{j,  m_j;~ k,  m_k}(\Omega_3) =\sum_{m_R=-j}^j \begin{pmatrix}
 C_{j, m_R; ~\frac{I}{2}, \frac{I}{2}}^{k, m_k} ~\Phi_{j,m_j;~j, m_R}(\Omega_3) \\
 C_{j, m_R; ~\frac{I}{2}, \frac{I}{2}-1}^{k, m_k} ~\Phi_{j,m_j;~j, m_R}(\Omega_3)\\
\vdots \\
 C_{j, m_R; ~\frac{I}{2}, -\frac{I}{2}}^{k, m_k} ~\Phi_{j,m_j;~j, m_R}(\Omega_3) 
\end{pmatrix}. \label{vectorlikespinsphereharmo}
\end{align}
\end{subequations}
Here, $d_{N+\frac{I}{2}+1, -j+k, j+k+1}$ in (\ref{exdefgmfunc}) stand for Wigner's small $D$-matrix (\ref{matrissywig}),   $C_{j, m_R; ~I/2, s_z}^{k, m_k}$ in (\ref{vectorlikespinsphereharmo}) represent the Clebsch-Gordan coefficients, and $\Phi_{j,~m_j;~j, m_R}(\Omega_3)$ denote  the $SO(4)$ spherical harmonics  \cite{Hasebe-2018}. 
 From (\ref{idenpanni}), the $SO(4)$ bi-spins (\ref{jkind}) now become   
\be
(j, k)_4\equiv (\frac{n}{2}+\frac{I}{4} +\frac{s}{2}, ~\frac{n}{2}+\frac{I}{4} -\frac{s}{2})_4, \label{formulajandk}
\ee
where 
\be
n=0, 1, 2, 3,  \cdots, N, ~~~~~~~~~s 
 =\frac{I}{2}, ~\frac{I}{2}-1, \cdots, -\frac{I}{2}. 
\label{jminuskcond}
\ee 
Equation (\ref{formulajandk}) implies that the Hilbert space of the $N$th $SO(5)$ Landau level consists of the smaller Hilbert spaces of the inner $SO(4)$ Landau levels: 
\be
\mathcal{H}_{SO(5)}^{(p=N+I,q=N)} = \bigoplus_{0\le n \le N} ~\bigoplus_{-\frac{I}{2} \le s \le {\frac{I}{2}}} \mathcal{H}_{SO(4)}^{[n,s]}.  \label{hilbertsumso45}
\ee
For instance, the LLL $(N=0)$ of $I=1$ holds  fourfold degeneracy made of two $SO(4)$ irreducible representations, $[n,s]=[0, 1/2]$ and $[0,-1/2]$,\footnote{The states of (\ref{lllfourbasis}) are essentially equal to those of (\ref{lowertwocomp}): 
\be
\bs{\psi}_1 =\frac{\sqrt{3}}{2\pi}\bs{\psi}_1^{[0,-\frac{1}{2}]},~~\bs{\psi}_1 =\frac{\sqrt{3}}{2\pi}\bs{\psi}_2^{[0,-\frac{1}{2}]},~~\bs{\psi}_3 =-\frac{\sqrt{3}}{2\pi}\bs{\psi}_3^{[0,-\frac{1}{2}]},~~\bs{\psi}_4 =-\frac{\sqrt{3}}{2\pi}\bs{\psi}_4^{[0,-\frac{1}{2}]}.  
\ee
} 
\begin{align}
&\bs{\psi}_1\equiv \bs{\psi}_{0; 1/2, 1/2; 0,0} =  
-\frac{\sqrt{3}}{2\pi} \sin\frac{\xi}{2}
\begin{pmatrix}
 \cos \chi-i\sin\chi \cos\theta \\
-i \sin\chi \sin\theta e^{i\phi} 
\end{pmatrix}, ~~~\bs{\psi}_2\equiv \bs{\psi}_{0; 1/2, -1/2; 0,0} = 
 \frac{\sqrt{3}}{2\pi} \sin\frac{\xi}{2}
\begin{pmatrix}
i \sin\chi \sin\theta  e^{-i\phi} \\
- \cos \chi-i\sin\chi \cos\theta
\end{pmatrix} , \nn\\
&\bs{\psi}_3\equiv \bs{\psi}_{0; 0, 0;  1/2, 1/2} 
=  
-\frac{\sqrt{3}}{2\pi}\begin{pmatrix}
\cos\frac{\xi}{2} \\
0 
\end{pmatrix} , ~~~\bs{\psi}_4\equiv \bs{\psi}_{0; 0, 0; 1/2, -1/2} =
  -\frac{\sqrt{3}}{2\pi}\begin{pmatrix}
0 \\
\cos\frac{\xi}{2}
\end{pmatrix}. \label{lllfourbasis}
\end{align}

\subsection{$SO(5)$  matrix generators for arbitrary irreducible representation}\label{subsec:arbitso5mat}

\begin{figure}[tbph]
\center
\includegraphics*[width=170mm]{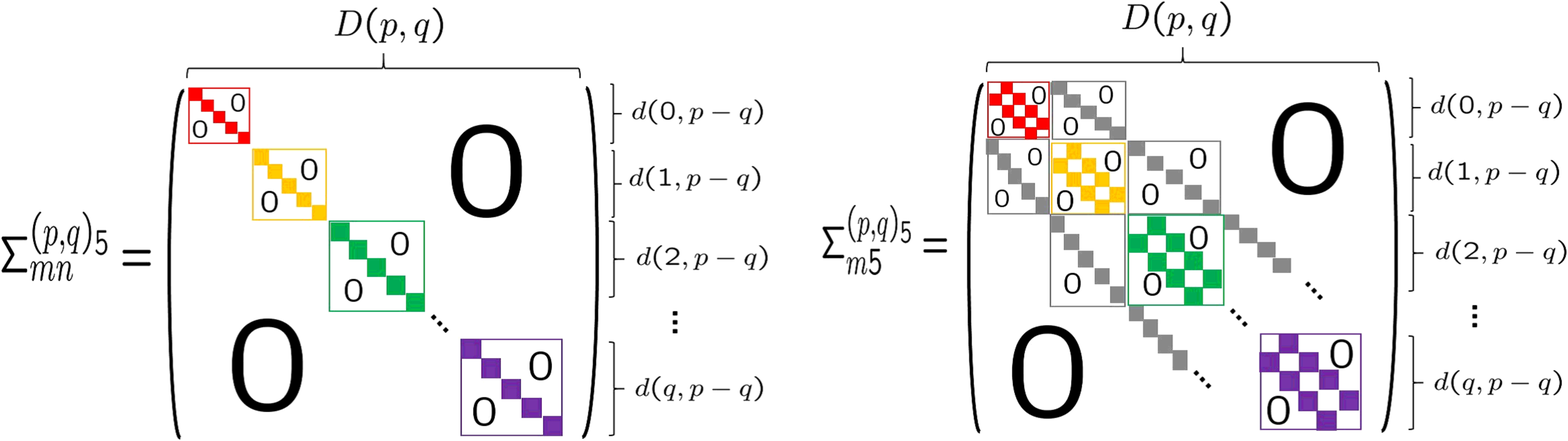}
\caption{General matrix form of the $SO(5)$ generators. The $SO(4)$ block matrices with non-zero  elements are denoted as the filled squares and rectangles.  }
\label{Sigmat.fig}
\end{figure}

We next investigate  the  matrix form of the $SO(5)$ generators of arbitrary irreducible representations. 
For notational brevity, with the understanding of (\ref{idenpanni}) we simply represent $\bs{\psi}_{N,; j, m_j; k, m_k}$ (\ref{x5so5monopolehamonicsnorm}) as 
\be 
\bs{\psi}_{\alpha}^{(p,q)_5} 
\ee 
where 
\be
\alpha =(j,m_j; k, m_k)=1,2,\cdots, D(p, q).
\ee
As the $SO(5)$ monopole harmonics realize a $(p,q)_5$ irreducible representation under the transformations generated by $L_{ab}$, 
\be
L_{ab}\bs{\psi}_{\alpha}^{(p,q)_5} =\bs{\psi}^{(p,q)_5}_{\beta}(\Sigma_{ab}^{(p,q)_5})_{\beta\alpha}, 
\ee
we can derive the $SO(5)$ matrix generators of $(p, q)_5$ by
\be
({\Sigma}^{(p,q)_5}_{a b})_{\alpha\beta} = 
 \int_{S^4}d\Omega_4 ~{\bs{\psi}^{(p,q)_5}_{\alpha}}^{\dagger}~L_{ab}~\bs{\psi}^{(p,q)_5}_{\beta}.  \label{expecsigmaab}
\ee
For instance  from (\ref{lllfourbasis}), 
$\Sigma_{ab}^{(1,0)_5}$  are derived as\footnote{With the $SO(5)$ gamma matrices 
\be
\gamma_m =\begin{pmatrix}
 0 & \bar{q}_m \\
q_m & 0 
\end{pmatrix}, ~~~\gamma_5=
\begin{pmatrix}
-1 & 0 \\
0 & 1
\end{pmatrix},  
\ee
Eqs.(\ref{sigmnm5}) are simply given by  
\be
\Sigma_{ab}^{(1,0)_5} =-i\frac{1}{4}[ \gamma_a, \gamma_b].
\ee
}  
\be
\Sigma_{mn}^{(1,0)_5} =\frac{1}{2}\begin{pmatrix}
\eta_{mn}^i\sigma_i & 0 \\
0 & \bar{\eta}_{mn}^i\sigma_i 
\end{pmatrix},
~~~
\Sigma_{m5}^{(1,0)_5} =i\frac{1}{2}\begin{pmatrix}
 0 & -\bar{q}_m \\
q_m & 0 
\end{pmatrix}  , \label{sigmnm5}
\ee
where $\eta_{mn}^i$ and $\bar{\eta}_{mn}^i$ are the 't Hooft symbols (\ref{defthooftsymbols}), and $q_m$ and $\bar{q}_m$ denote the quaternions and their quaternion conjugates:  
\be
q_m =\{-i\sigma_i, 1\}, ~~\bar{q}_m =\{i\sigma_i, 1\}. \label{matquaternions}
\ee
The $SO(4)$ decomposition  (\ref{hilbertsumso45}) implies 
\be
{\Sigma}_{m n}^{(p,q)_5}  =\bigoplus_{0\le n \le q} \bigoplus_{-\frac{p-q}{2}\le s \le \frac{p-q}{2}} ~\sigma_{m n}^{(j=\frac{n}{2}+\frac{p-q}{4}+\frac{s}{2}, k= \frac{n}{2}+\frac{p-q}{4}-\frac{s}{2})_4} ,   
\label{sigmahatsigmas}
\ee
where  $\sigma_{m n}^{(j,k)_4}$ are the $SO(4)\simeq SU(2)_L\otimes SU(2)_R$ matrix generators with index $(j, k)_4$,   
\be
\sigma_{m n}^{(j,k)_4}\equiv \eta_{m n}^i S_i^{(j)}\otimes \bs{1}_{2k+1} + \bs{1}_{2j+1}\otimes \bar{\eta}_{m n}^i S_i^{(k)}. \label{so4irrepgenemat}
\ee
More specifically, 
\be
\Sigma_{mn}^{(p,q)_5} \equiv 
\begin{pmatrix}
\sigma_{mn}^{[n=0]} & 0 & 0 & 0 & 0\\
0 &  \sigma_{mn}^{[n=1]} & 0 & 0 & 0 \\
0 & 0 &  \sigma_{mn}^{[n=2]} & 0 & 0 \\ 
0 & 0 &  0 & \ddots   & 0 \\ 
0 & 0 &  0 & 0  & \sigma_{mn}^{[n=q]}
\end{pmatrix}, \label{directsumsigmapre}
\ee
where  $\sigma_{mn}^{[n]}$ denotes the $d(n, p-q)\times d(n, p-q)$ square matrix that is 
further block-diagonalized:    
\be
\sigma_{mn}^{[n]} \equiv 
\begin{pmatrix}
\sigma_{mn}^{(\frac{p-q}{2} +\frac{n}{2}, \frac{n}{2})_4} & 0 & 0 & 0 \\
0 & \sigma_{mn}^{(\frac{p-q}{2}+\frac{n}{2}-\frac{1}{2}, \frac{n}{2}+\frac{1}{2})_4} & 0 & 0 \\
0 & 0 & \ddots & 0 \\
0 & 0 & 0 & \sigma_{mn}^{(\frac{n}{2}, \frac{n}{2}+\frac{p-q}{2} )_4}
\end{pmatrix}. \label{blochdiagso4op}
\ee
See the left of Fig.\ref{Sigmat.fig}. 
Since 
 $L_{m 5}$ 
behave as an $SO(4)$  vector of the $SO(4)$ bi-spins,  
\be
(j,k)_4=(\frac{1}{2}, \frac{1}{2})_4, 
\ee
 the  $SU(2)$ selection rule indicates that the  matrix elements of ${L}_{m 5}$ take non-zero values only for  
\be
(\Delta j, \Delta k)_4 =(\frac{1}{2}, \frac{1}{2})_4, ~~(-\frac{1}{2}, \frac{1}{2})_4, ~~(\frac{1}{2}, -\frac{1}{2})_4, ~~(-\frac{1}{2}, -\frac{1}{2})_4.  
\ee
In other words, $\Sigma_{m5}^{(p,q)}$ have finite matrix elements only between  nearest $SO(4)$ irreducible representations in Fig.\ref{so5diagram.fig},  and the matrix form of the $\Sigma_{m5}^{(p,q)}$ is depicted at  
 the right of Fig.\ref{Sigmat.fig}. The matrices  (\ref{sigmnm5}) actually fit the general matrix form of Fig.\ref{Sigmat.fig}.  
It should be emphasized  that while we used Yang's monopole harmonics, the obtained $SO(5)$ matrix generators do $\it{not}$ depend on the functional forms  specific to Yang's monopole harmonics and are universal for any $SO(5)$ irreducible representations. 

\section{$SO(5)$ monopole harmonics as non-linear realization}\label{sec:so5monoharnl}

Here, we discuss  how the non-linear realization is related to  quantum mechanics with gauge symmetry.   While we focus on  the $SO(5)$ case, the obtained results can easily be  generalized to arbitrary groups.  

\subsection{$SO(5)$ non-linear realization and $SO(4)$ gauge symmetry}\label{subsec:formalnonline}

Let us  consider the non-linear realization of the $SO(5)$ group for the coset manifold 
\be 
S^4\simeq SO(5)/SO(4). 
\ee 
In the context of quantum field theory, the coset represents the field manifold  associated with the spontaneous symmetry breaking of  
$SO(5)~\rightarrow~SO(4).$ 
With the broken generators 
\be
\Sigma_{m 5}^{(p,q)_5}, ~~~(m=1,2,3,4)
\ee
we can construct the associated non-linear realization matrix 
\be
\Psi^{(p,q)_5}(\Omega_4)=e^{i\sum_{m=1}^4 \alpha_m(\Omega_4) \Sigma_{m5}^{(p,q)_5}}, \label{genepsiso5s4}
\ee
where  $\alpha_m$ are parameters to be determined. With an element of the unbroken $SO(4)$ group,  
\be
H=e^{\frac{1}{2}\sum_{m,n=1}^4 \omega_{mn}\Sigma^{(p,q)_5}_{mn}}, \label{hexp}
\ee
 the $SO(5)$ group element is  locally represented as 
\be
H^{\dagger} \cdot \Psi^{(p,q)_5}.
\ee
Equation (\ref{directsumsigmapre}) implies  that $H$ (\ref{hexp}) is expressed 
as  a completely reducible representation of the $SO(4)$: 
\be
H=\begin{pmatrix}
h^{[0]} & 0 & 0 & 0  \\
0 & h^{[1]} & 0 & 0  \\
0 & 0 &  \ddots & 0  \\
 0 & 0 & 0 &  h^{[q]}
\end{pmatrix}=\bigoplus_{n=0}^q h^{[n]}, 
\ee
and each of the block matrices is further block-diagonalized: 
\be
h^{[n]} =\begin{pmatrix}
h^{[{n} , \frac{p-q}{2}]} & 0 & 0 & 0  \\
0 & h^{[n, \frac{p-q}{2} -1]} & 0 & 0  \\
0 & 0 &  \ddots & 0  \\
 0 & 0 & 0 &  h^{[n, -\frac{p-q}{2} ]}
\end{pmatrix} =\bigoplus_{-\frac{p-q}{2} \le s \le  \frac{p-q}{2}}  h^{[n,s]}. \label{subblockshs}
\ee
Recall that $[n,s]$ specifies the $SO(4)$ bi-spin indices (\ref{jknsinter}). 
Assume that the unbroken $SO(4)$ transformation acts as  a ``gauge'' transformation\footnote{In the context of field theory, Eq.(\ref{gaugetranspsi}) is called the hidden local symmetry of non-linear realization. }
\be
\Psi^{(p,q)_5} ~~\rightarrow~~H^{\dagger}\cdot \Psi^{(p,q)_5}, \label{gaugetranspsi}
\ee
while the global transformation $G\in SO(5)$ acts as a right action: 
\be
\Psi^{(p,q)_5} ~~\rightarrow ~~\Psi^{(p,q)_5} \cdot G. \label{psiso5tra}
\ee
The corresponding connection is introduced as 
\be
\mathcal{A}_{a} =-i\Psi^{(p,q)_5}\partial_a {\Psi^{(p,q)_5}}^{\dagger}=\begin{pmatrix}
A^{[0]}_{a} & A^{[0,1]}_{a}  & \cdots   & A^{[0,q]}_{a}   \\
A^{[1,0]}_{a}  & A^{[1]}_{a}  & \cdots  & A^{[1,q]}_{a}   \\
\vdots   & \vdots &  \ddots & \vdots   \\
 A^{[q,0]}_{a}  & A^{[q,1]}_{a}  & \cdots &  A^{[q]}_{a} 
\end{pmatrix}.  \label{mathprea}
\ee
Under the transformation (\ref{gaugetranspsi}),  (\ref{mathprea}) transforms as an $SO(4)$ gauge field as anticipated:  
\be
\mathcal{A}_a ~~\rightarrow~~H^{\dagger}\mathcal{A}_a H-iH^{\dagger}\partial_a H.  \label{watransgauge}
\ee
However, note that $\mathcal{A}_a$ (\ref{mathprea}) is a pure gauge whose curvature identically vanishes. 
To realize a physical gauge field, we utilize 
the block-diagonal parts of 
(\ref{mathprea}), 
\be
A_a \equiv \begin{pmatrix}
A^{[0]}_{a} & 0  & 0   & 0   \\
0 & A^{[1]}_{a}  & 0 &0  \\
0   & 0 &  \ddots & 0   \\
0 & 0 & 0 &  A^{[q]}_{a} 
\end{pmatrix}=\bigoplus_{n=0}^q A_a^{[n]},   
\ee
and each of the  block matrices is given by 
\be
A_a^{[n]} =\begin{pmatrix}
A_a^{[{n} , \frac{p-q}{2}]} & 0 & 0 & 0  \\
0 & A_a^{[n , \frac{p-q}{2}-1]} & 0 & 0  \\
0 & 0 &  \ddots & 0  \\
 0 & 0 & 0 &  A_a^{[n, -\frac{p-q}{2} ]}
\end{pmatrix} =\bigoplus_{-\frac{p-q}{2}\le s \le  \frac{p-q}{2}}  A_a^{[n,s]}. \label{furtherblockagau}
\ee
Under the transformation (\ref{gaugetranspsi}), $A_a$ transforms similarly to (\ref{watransgauge}):  
\be
A_a ~~\rightarrow~~H^{\dagger}A_a H-iH^{\dagger}\partial_a H. \label{aatransgauge}
\ee
We see that $A_a$ is no longer a pure gauge field in the sense that the corresponding curvature, $F_{ab} =\partial_a A_b-\partial_b A_a +i[A_a, A_b]$, does not vanish. It is also obvious that $A_a$ are invariant under the global $SO(5)$ transformation (\ref{psiso5tra}).  
With the  $A_a$, we can introduce the covariant derivatives and angular momentum operators for the non-linear representation as\footnote{Under the gauge and the global transformations, the quantities defined by (\ref{defdajab}) respectively transform as
\be 
D_a \Psi^{(p,q)_5} ~~\rightarrow~~H^{\dagger}\cdot D_a\Psi^{(p,q)_5}, ~~~~~~~~J_{ab}\Psi^{(p,q)_5}~~\rightarrow~~H^{\dagger}\cdot J_{ab}\Psi^{(p,q)_5}, \label{deddjab}
\ee
and 
\be 
D_a \Psi^{(p,q)_5} ~~\rightarrow~~D_a\Psi^{(p,q)_5} \cdot G,~~~~~~~~J_{ab}\Psi^{(p,q)_5}~~\rightarrow~~J_{ab}\Psi^{(p,q)_5}\cdot G. 
\ee
} 
\be
D_a \Psi^{(p,q)_5} \equiv \partial_a \Psi^{(p,q)_5} +iA_a \Psi^{(p,q)_5}, ~~~~~J_{ab} \Psi^{(p,q)_5} \equiv (-ix_aD_b+ix_bD_a +r^2 F_{ab})\Psi^{(p,q)_5}. \label{defdajab}
\ee 

Let us focus on  the smaller $SO(4)$ gauge transformations denoted by $h^{[n,s]}$ of (\ref{subblockshs}) that carry the $SO(4)$ bi-spin indices:  
\be
(j,k)_4= (\frac{n}{2}+\frac{p-q}{4}+\frac{s}{2}, ~\frac{n}{2}+\frac{p-q}{4}-\frac{s}{2}  )_4. \label{so4indnsjk}
\ee
We represent $\Psi$ (\ref{genepsiso5s4}) as 
\be
\Psi^{(p,q)_5} =\begin{pmatrix}
{\Psi}_1^{[0]} & {\Psi}_2^{[0]} & \cdots & {\Psi}_{D(p,q)}^{[0]} \\
{\Psi}_1^{[1]} & {\Psi}_2^{[1]} & \cdots & {\Psi}_{D(p, q)}^{[1]} \\
\vdots & \vdots & \ddots & \vdots  \\
{\Psi}_1^{[q]} & {\Psi}_2^{[q]} & \cdots & {\Psi}_{D(p, q)}^{[q]}
\end{pmatrix},  
\ee
and each block ${\Psi}_{(\alpha=1,2,\cdots, D(p,q))}^{[n]}$ ~$(n=0,1,2,\cdots, q)$ which we call the $n$-sector of  $\Psi^{(p,q)_5} $ takes the form of 
\be
{\Psi}_\alpha^{[n]} =
\begin{pmatrix}
\bs{\psi}_\alpha^{[n, \frac{p-q}{2}]} \\
\bs{\psi}_\alpha^{[n, \frac{p-q}{2}-1]} \\
\vdots \\
\bs{\psi}_\alpha^{[n, s]}  
\\ \vdots \\
\bs{\psi}_\alpha^{[n, -\frac{p-q}{2}]}
\end{pmatrix}. \label{psicolumori}
\ee
The gauge (\ref{gaugetranspsi}) and the global transformations (\ref{psiso5tra}), respectively, act to the  $\bs{\psi}_\alpha^{[n, s]}$ $(-\frac{p-q}{2} \le s \le \frac{p-q}{2})$  as  
\be
\bs{\psi}^{[n,s]}_\alpha ~~\rightarrow~~{h^{[n,s]}}^{\dagger}\bs{\psi}^{[n,s]}_\alpha,~~~~~~~~~~~~\bs{\psi}^{[n,s]}_\alpha ~~\rightarrow~~\sum_{\beta=1}^{D(p,q)}\bs{\psi}^{[n,s]}_\beta ~G_{\beta\alpha}. \label{transformpsisubsub}
\ee
The gauge field $A_a^{[n,s]}$ in  (\ref{furtherblockagau}) is represented as  
\be
A_a^{[n,s]} =-i \sum_{\alpha=1}^{D(p,q)} \bs{\psi}_\alpha^{[n,s]} \partial_a {\bs{\psi}_\alpha^{[n,s]}}^{\dagger}, 
\label{so4connectionassocpi}
\ee
which transforms as 
\be
A_a^{[n,s]} ~~\rightarrow~~{h^{[n,s]}}^{\dagger}A_a^{[n,s]} h^{[n,s]}-i{h^{[n,s]}}^{\dagger}\partial_a h^{[n,s]}.  \label{aatransgaugen}
\ee
Using (\ref{so4connectionassocpi}), we can construct the covariant derivatives and the angular momentum operators  as 
\begin{align}
&D_a^{[n,s]}\bs{\psi}^{[n,s]}_\alpha =\partial_a\bs{\psi}^{[n,s]}_\alpha +iA_{a}^{[n,s]}\bs{\psi}^{[n,s]}_\alpha, \nn\\
&J_{ab}^{[n,s]}\bs{\psi}^{[n,s]}_\alpha \equiv (-ix_a D_b^{[n,s]} +ix_bD_a^{[n,s]} +r^2F^{[n,s]}_{ab})\bs{\psi}^{[n,s]}_\alpha.   
\end{align}
The second equation of (\ref{transformpsisubsub}) implies that  the set $\bs{\psi}_{\alpha=1,2,\cdots, D(p,q)}^{[n,s]}$ constitutes  an $SO(5)$ irreducible representation with $(p,q)_5$, and at the same time,  $\bs{\psi}_\alpha^{[n,s]}$ enjoys the $SO(4)$ gauge symmetry 
of the $SO(4)$ bi-spin indices (\ref{so4indnsjk}). 
The physical quantities that hold such features are nothing but the $SO(5)$ monopole harmonics.  

\subsection{Determination of the $SO(5)$ non-linear realization}

Our next task is to determine the parameters $\alpha_m$ of  
the non-linear realization (\ref{genepsiso5s4}). 
For this purpose, it is sufficient to    consider 
the simplest case 
$\Sigma_{ab}^{(1,0)_5}$ (\ref{sigmnm5}), in which the non-linear realization (\ref{genepsiso5s4}) reads 
\be
\Psi^{(1,0)_5}(\Omega_4)=  \begin{pmatrix}
\cos (\frac{\alpha}{2})1_2 & \sin (\frac{\alpha}{2}) ~ \frac{1}{\alpha}{\alpha}_m \bar{q}_m \\
- \sin( \frac{\alpha}{2}) ~\frac{1}{\alpha}\alpha_m {q}_m & \cos (\frac{\alpha}{2}) 1_2
\end{pmatrix}  \label{simplestnonlin}
\ee
with $\alpha \equiv \sqrt{{\alpha_m}^2}$. 
According to the discussions of Sec.\ref{subsec:formalnonline},  
we rewrite  (\ref{simplestnonlin}) in the following form 
\be
\Psi^{(1,0)_5} (\Omega) =
\begin{pmatrix}
\bs{\psi}_1^{[0, \frac{1}{2}]} & \bs{\psi}_2^{[0, \frac{1}{2}]} & \bs{\psi}_3^{[0, \frac{1}{2}]} & \bs{\psi}_4^{[0, \frac{1}{2}]} \\
\bs{\psi}_1^{[0, -\frac{1}{2}]} & \bs{\psi}_2^{[0, -\frac{1}{2}]} & \bs{\psi}_3^{[0, -\frac{1}{2}]} & \bs{\psi}_4^{[0, -\frac{1}{2}]} 
\end{pmatrix} \label{comppsisimple}
\ee
to see that 
the set of the upper and lower two columns, respectively, represents the monopole harmonics of $(p, q)_5=(1,0)_5$  in the $SU(2)$ monopole background and in the $SU(2)$ anti-monopole background.   
Recall the (anti-)monopole harmonics   (\ref{lllfourbasis}) to   construct 
\be
\frac{2\pi} {\sqrt{3}}~ \begin{pmatrix} \bs{\psi}_{1}  & \bs{\psi}_{2}  & -\bs{\psi}_{3}  & -\bs{\psi}_{4}\end{pmatrix} = \frac{1}{\sqrt{2(1+x_5)}} \begin{pmatrix}
-x_m q_m &  (1+x_5) 1_2\end{pmatrix} , 
\ee
which should be identified as the lower two columns of (\ref{simplestnonlin}). 
Now $\alpha_m$ can be identified as   
\be
\alpha_m(\Omega_4) =\xi ~y_m, 
\ee
where $y_{m}$ $(m=1,2,3,4)$ denote the coordinates on the hyper-latitude at the azimuthal angle $\xi$ on $S^4$: 
\be
y_m \equiv \frac{1}{\sin\xi}~x_m =\{\sin\chi\sin\theta\cos\phi, \sin\chi\sin\theta\sin\phi, \sin\chi\cos\theta, \cos\chi\}~\in ~S^3.
\ee
The non-linear realization (\ref{simplestnonlin}) is  represented as  
\be
\Psi^{(1,0)_5}(\Omega_4) =\frac{1}{\sqrt{2(1+x_5)}} 
\begin{pmatrix}
(1+x_5)1_2 & x_{m}\bar{q}_{m} \\
-x_{m }q_{m} & (1+x_5)1_2
\end{pmatrix}. \label{exp01psi}
\ee
For general representation $(p,q)_5$,   the non-linear realization is given by 
\be
\Psi^{(p,q)_5}(\Omega_4) =e^{i\xi \sum_{m=1}^4 y_m\Sigma_{m 5}^{(p,q)_5}}, \label{finalpsimatso5}
\ee
which naturally generalizes the  $SO(3)$ case (\ref{cosetrepsu2}).   
It is straightforward to check that 
(\ref{finalpsimatso5}) covariantly transforms under the $SO(5)$ rotations generated by $J_{ab}$ (\ref{defdajab}), 
\be
J_{ab} \Psi^{(p,q)_5}(\Omega_4) =\Psi^{(p,q)_5}(\Omega_4) ~\Sigma_{ab}^{(p,q)_5}, \label{so5covpsimat}
\ee
which  implies   
\be
\sum_{a<b}{J_{ab}}^2 ~\Psi^{(p,q)_5}(\Omega_4) = \Psi^{(p,q)_5}(\Omega_4) ~\sum_{a<b} {\Sigma_{ab}^{(p,q)_5}}^2 =\lambda(p,q) \Psi^{(p,q)_5}(\Omega_4).  \label{caseigenvalpsimat}
\ee
In the language of $\bs{\psi}_{\alpha}^{[n,s]}$,  Eq.(\ref{caseigenvalpsimat}) is translated as 
\be
\sum_{a<b}{J_{ab}^{[n,s]}}^2 \bs{\psi}^{[n,s]}_\alpha =\lambda(p,q) ~\bs{\psi}^{[n,s]}_\alpha . \label{smallpsicasim} 
\ee
Note that (\ref{smallpsicasim}) signifies  that $\bs{\psi}^{[n,s]}_{\alpha}$ are the $SO(5)$ monopole harmonics  with the eigenvalue value $\lambda(p,q)$ in the   $SO(4)$ monopole background with  $(\frac{I_+}{2}, \frac{I_-}{2})_4 = (\frac{n}{2}+\frac{p-q}{4}+\frac{s}{2}, \frac{n}{2}+\frac{p-q}{4}-\frac{s}{2})_4$.

\section{$SO(5)$ Landau problem in the $SO(4)$ monopole background}\label{sec:so5landauinso4}

We now apply  the techniques of the non-linear realization  to the $SO(5)$ Landau problem in the $SO(4)$ monopole background. 
In the context of the Landau model,   $p$ and $q$ are quantities to be determined.   

\subsection{The $SO(4)$ monopole and  $SO(5)$ Landau Hamiltonian}

Before proceeding to the $SO(5)$ Landau problem, 
we explain topological features  of  the $SO(4)$ monopole gauge field. The $SO(4)$ monopole is simply introduced with replacement of the $SU(2)$ spin matrices of the Yang monopole (\ref{zhanghusu2gauge}) with the $SO(4)$ bi-spin matrices:  
\be
A_{m} =-\frac{1}{r(r+x_5)}\sigma_{m n}^{(\frac{I_+}{2}, \frac{I_-}{2})_4}x_{n},~~~A_{5}=0, \label{so4monoinput}
\ee
where  
\be
\sigma_{mn}^{(\frac{I_+}{2}, \frac{I_-}{2})_4} =\eta_{mn}^iS_i^{(\frac{I_+}{2})}\otimes \bs{1}_{I_-+1} + \bs{1}_{I_++1} \otimes \bar{\eta}_{mn}^iS_i^{(\frac{I_-}{2})}. 
\ee
The $SO(4)$ monopole is conformally equivalent to the $SO(4)$ instanton on $\mathbb{R}^4$ that is a solution of the pure Yang-Mills field equations  \cite{BPST-1975,Jackiw-Rebbi-1976,Hasebe-2020-3}.   
The $SO(4)$ monopole gauge field (\ref{so4monoinput}) can be expressed as 
\be
A=A_a dx_a =A^{(+)} \oplus \bs{1}_{I_-+1} + \bs{1}_{I_+ +1} \otimes A^{(-)},
\ee
where $A^{(+)}$ and $A^{(-)}$ denote the $SU(2)$ monopole  field and  the $SU(2)$ anti-monopole  field, respectively: 
\be
A^{(+)}= -\frac{1}{r(r+x_5)}\eta_{m n}^i S_i^{(\frac{I_+}{2})} x_{n}dx_m, ~~~~~ A^{(-)} =-\frac{1}{r(r+x_5)}\bar{\eta}_{m n}^i S_i^{(\frac{I_-}{2})} x_{n}dx_m.
\ee
The corresponding field strength, $F_{ab}=\partial_a A_b -\partial_b A_a +i [A_a, A_b]$, is derived by 
\be
F_{mn} =-\frac{1}{r^2}x_m A_n +\frac{1}{r^2}x_n A_m +\frac{1}{r^2}\sigma_{mn}^{(\frac{I_+}{2}, \frac{I_-}{2})_4}, ~~~F_{m 5} =-F_{5 m} =\frac{1}{r^2}(r+x_5) A_m, \label{fmnso4}
\ee
which satisfy 
\be
\sum_{a<b}{F_{ab}}^2 =\frac{1}{r^4}\sum_{m<n} {\sigma_{mn}^{(\frac{I_+}{2}, \frac{I_-}{2})_4}}^2=\frac{1}{2r^4}(I_+(I_++2) +I_-(I_-+2)) \bs{1}_{(I_+ +1)(I_- +1)} . \label{so4fiesqu}
\ee 
With the vierbein $e^m$ of $S^4$,  (\ref{fmnso4}) can be concisely expressed as  
\be
F =\frac{1}{2}F_{ab}~dx_a \wedge dx_b =\frac{1}{2} e^{m}\wedge e^{n}~\sigma_{mn}^{(\frac{I_+}{2}, \frac{I_-}{2})_4}.
\ee
The $SO(4)$ group hosts two invariant tensors, $i.e.$, Kronecker delta symbol and  Levi-Civita  four-rank tensor, which allow us to introduce two  $SO(4)$ gauge invariant topological invariants \cite{Eguchi-Freund-1976}, the (total) second  Chern number and a generalized Euler number  
(see Appendix \ref{appendix:so4monopole} for details):
\begin{subequations}
\begin{align}
&c_2\equiv \frac{1}{8\pi^2}\int \tr(F^2)=\frac{1}{8\pi^2}\int\tr(\mathcal{F}^2)  =\frac{1}{32\pi^2}\int F^{m_1 m_2} F^{m_3 m_4}~\tr(\sigma_{m_1m_2} \sigma_{m_3m_4}), \label{chernumbsec}\\
&\tilde{c}_2\equiv  \frac{1}{8\pi^2}\int \tr(F\mathcal{F}) = \frac{1}{8\pi^2}\int \tr(\mathcal{F} F) = \frac{1}{64\pi^2}\int \epsilon^{m_3m_4m_5m_6}F^{m_1 m_2} F^{m_3 m_4}~\tr(\sigma_{m_1m_2} \sigma_{m_5m_6}), \label{genereul}
\end{align}\label{twotopinv}
\end{subequations}
where  
\be
F \equiv \frac{1}{2}F^{m_1 m_2}\sigma_{m_1 m_2}, ~~~~~~~\mathcal{F}\equiv \frac{1}{4}\epsilon^{m_1m_2m_3m_4}F_{m_1m_2}\sigma_{m_3 m_4}. 
\ee
For $F_{mn} =e_{m}\wedge e_{n}$ and $\sigma_{mn}=\sigma_{mn}^{(\frac{I_+}{2}, \frac{I_-}{2})_4}$, 
  Eq.(\ref{twotopinv}) is evaluated as\footnote{
For  $(j,k)=(1/2, 0), (0, 1/2), (1/2, 0)\oplus (0, 1/2), (1/2, 1/2)$, the $SO(4)$ matrix generators are respectively given by  
\be
\sigma_{mn} =\frac{1}{2}\eta_{mn}^i\sigma_i,~\frac{1}{2}\bar{\eta}_{mn}^i\sigma_i,~\frac{1}{2}
\begin{pmatrix}
\eta_{mn}^i\sigma_i & 0 \\
0 & \bar{\eta}_{mn}^i\sigma_i
\end{pmatrix}, ~~\frac{1}{2}\eta_{mn}^i\sigma_i\otimes 1_2+1_2\oplus \frac{1}{2}\bar{\eta}_{mn}^i\sigma_i, 
\ee
and the  topological invariants (\ref{twotopinv}) are evaluated as  
\be
(c_2, \tilde{c}_2)=(1, 1),~(-1, 1),~(0, 2),~(0,4).
\ee
In deriving (\ref{cchimonop}), we used the formula 
\be
\tr(\sigma_{m_1 m_2}^{(j,k)_4}\sigma_{m_3 m_4}^{(j,k)_4}) 
= \frac{(2j+1)(2k+1)}{3} \biggl((j(j+1)+k(k+1))(\delta_{m_1m_3}\delta_{m_2m_4}-\delta_{m_1 m_4}\delta_{m_2 m_3}) + (j(j+1)-k(k+1))\epsilon_{m_1m_2m_3m_4} \biggr). 
\ee
} 
\begin{subequations}
\begin{align}
&c_2^{(\frac{I_+}{2} ,\frac{I_-}{2})}=  \frac{1}{6}(I_++1)(I_-+1)(I_+(I_++2)-I_-(I_-+2)), \label{c2=g} \\
&\tilde{c}_2^{(\frac{I_+}{2} ,\frac{I_-}{2})}= \frac{1}{6}(I_++1)(I_-+1)(I_+(I_++2)+I_-(I_-+2)).  \label{chiexp}
\end{align}\label{cchimonop}
\end{subequations}
Meanwhile, from the homotopy theorem 
\be
\pi_3(SO(4)) 
\simeq \pi_3(SU(2))\oplus \pi_3(SU(2)) \simeq \mathbb{Z}\oplus \mathbb{Z},  
\ee
we can  introduce two distinct second Chern numbers corresponding to the monopole and the anti-monopole, 
\begin{subequations}
\begin{align}
&c_2^+ =\frac{1}{2(2\pi)^2}\int_{S^4}\tr(F_+^2) =+\frac{1}{6}I_+(I_+ +1)(I_+  +2) ,\\
&c_2^- =\frac{1}{2(2\pi)^2}\int_{S^4}\tr(F_-^2) =-\frac{1}{6}I_-(I_- +1)(I_-  +2),  
\end{align}
\end{subequations}
which are related to 
$c_2$ (\ref{c2=g}) and $\tilde{c}_2$ (\ref{chiexp})  as 
\be 
c_2^{(\frac{I_+}{2} ,\frac{I_-}{2})}=c_2^+ (I_-+1)+c_2^{-}(I_++1), ~~~~~
\tilde{c}_2^{(\frac{I_+}{2} ,\frac{I_-}{2})} =c_2^+ (I_-+1)-c_2^{-}(I_++1).
\ee 
The second Chern number 
$c_2$ essentially represents the sum of the two monopole charges, while the generalized Euler number $\tilde{c}_2$ represents their difference. They  may be reminiscent of  the topological invariants of  ($S_z$ conserved) quantum spin Hall effect \cite{Kane-Mele-2005, Bernevig-Zhang-2005, Sheng-Weng-Sheng-Haldane-2006}; the sum of two  Chern number signifies  quantized charge Hall conductance, while their difference  indicates  quantized spin Hall conductance.  
In the non-chiral case ${I_+} =I_-=\frac{I}{2}$ $(I=0,2,4,6,\cdots)$, though the second Chern number is trivial, the generalized Euler number is finite, 
\be
c_2^{(\frac{I}{4} ,\frac{I}{4})}=0, ~~~~~ \tilde{c}_2^{(\frac{I}{4}, \frac{I}{4})} = \frac{1}{48}I(I+2)^2(I+4), 
\ee
and $\tilde{c}_2$ is the unique topological quantity of the system. 

Replacing the $SU(2)$ gauge field with the $SO(4)$ gauge field,  we introduce the $SO(5)$  angular momentum operators in the $SO(4)$ monopole background in a similar manner to Sec.\ref{subsec:so5ori}: 
\be
L_{m n} =L_{m n}^{(0)} +\sigma_{m n}^{(\frac{I_+}{2}, \frac{I_-}{2})_4}, ~~~~L_{m 5} =-L_{5 m} =L_{m 5}^{(0)} -\frac{1}{r+x_5}\sigma_{m n}^{(\frac{I_+}{2}, \frac{I_-}{2})_4}x_{n}. 
\ee 
With covariant angular momentum operators $\Lambda_{ab} =-ix_aD_b+ix_bD_a$, we construct 
the $SO(5)$ Landau Hamiltonian in the $SO(4)$ monopole background: 
\begin{align}
H&=-\frac{1}{2M}(\partial_a +iA_a)^2\biggr|_{r=1} =\frac{1}{2M}\sum_{a<b}{\Lambda_{ab}}^2 
= \frac{1}{2M}(\sum_{a<b}{L_{ab}}^2  
- \sum_{a<b}{F_{ab}}^2) \nn\\
&=  \frac{1}{2M}\biggl(\sum_{a<b}{L_{ab}}^2 - \frac{1}{2}({I_+}({I_+}+2)+{I_-}({I_-}+2))
\biggr), \label{so5so4hamilandau}
\end{align}
and hence the energy eigenvalues of (\ref{so5so4hamilandau}) are expressed as  
\be
E=\frac{1}{2}\biggl(\lambda(p,q) -\frac{1}{2}(I_+(I_++2) +I_-(I_-+2))\biggr). \label{so5so4monolls}
\ee
Since the gauge field was introduced as an external gauge field that does not change its sign under the time-reversal transformation,   
the Landau Hamiltonian (\ref{so5so4hamilandau}) does not respect the time-reversal symmetry even in the non-chiral case. 

\subsection{$SO(5)$ Landau level eigenstates}

Let us first address how the $SO(5)$ Landau level eigenstates can be  identified as the non-linear realization.  
As discussed in Sec.\ref{subsec:formalnonline}, $\bs{\psi}_{\alpha=1,2,\cdots, D(p,q)}^{[n, s]}$ enjoy  the $SO(4)$ gauge symmetry 
with the $SO(4)$ bi-spin indices $(\ref{so4indnsjk})$, 
which in the context of the Landau model are identified with the $SO(4)$ monopole indices, 
\be
(\frac{I_+}{2}, \frac{I_-}{2})_4 = (\frac{n}{2} +\frac{p-q}{4}+\frac{s}{2}, \frac{n}{2} +\frac{p-q}{4}-\frac{s}{2})_4. \label{ipimfrompqs}
\ee
Since $n$ runs from $0$ to $q$, 
$q$ should be greater than  or equal to  $n$,\footnote{Recall the similar discussions in the $SO(3)$ Landau model around   (\ref{nl-g}). Equation (\ref{defllfromn}) is a generalization of (\ref{nl-g}).} so  we can define non-negative integers $N$ for $\it{each}$ $n$,  
\be
N \equiv q-n = 0, 1, 2, \cdots. \label{defllfromn}
\ee
The non-negative integer $N$ indicates the Landau level index in the $n$-sector, and then  $\bs{\psi}_{\alpha=1,2,\cdots, D(p,q)}^{[n, s]}$ represent the $N$th  Landau level eigenstates of the $n$-sector. We will discuss the energy levels in Sec.\ref{subsec:so5levels}.

\begin{figure}[tbph]
\center
\includegraphics*[width=170mm]{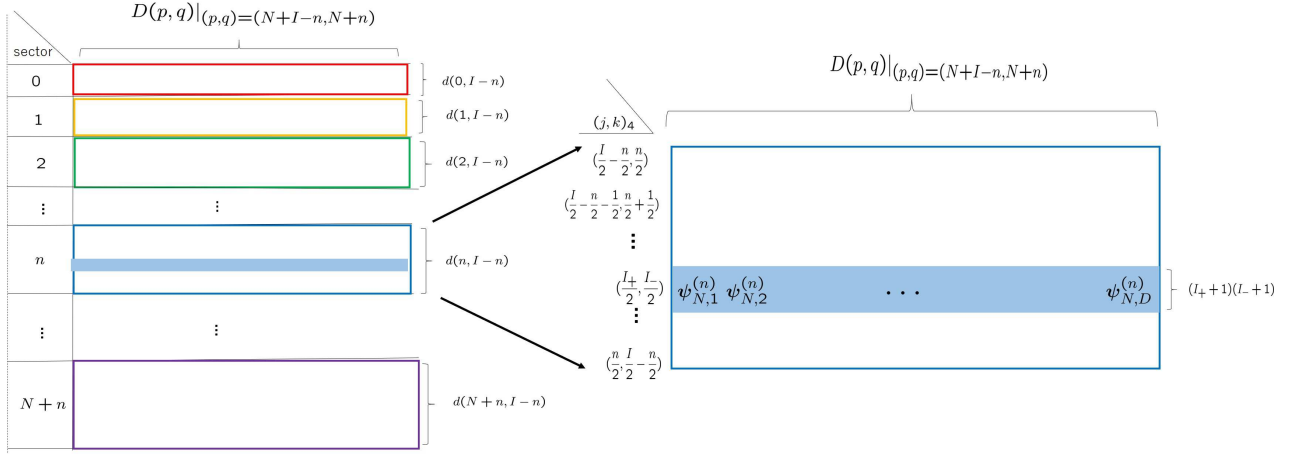}
\caption{ 
The $N$th Landau level eigenstates in the $n$-sector, $\bs{\psi}_{N, 1}^{(n)}, \bs{\psi}_{N, 2}^{(n)}, \cdots, \bs{\psi}_{N,D}^{(n)}$, can be found as the  block matrix  (the blue shaded region) in the $n$-sector of  $\Psi^{(p,q)}|_{p=N+I-n, ~q=N+n}$. }
\label{geneig.fig}
\end{figure}

\begin{figure}[tbph]
\center
\includegraphics*[width=160mm]{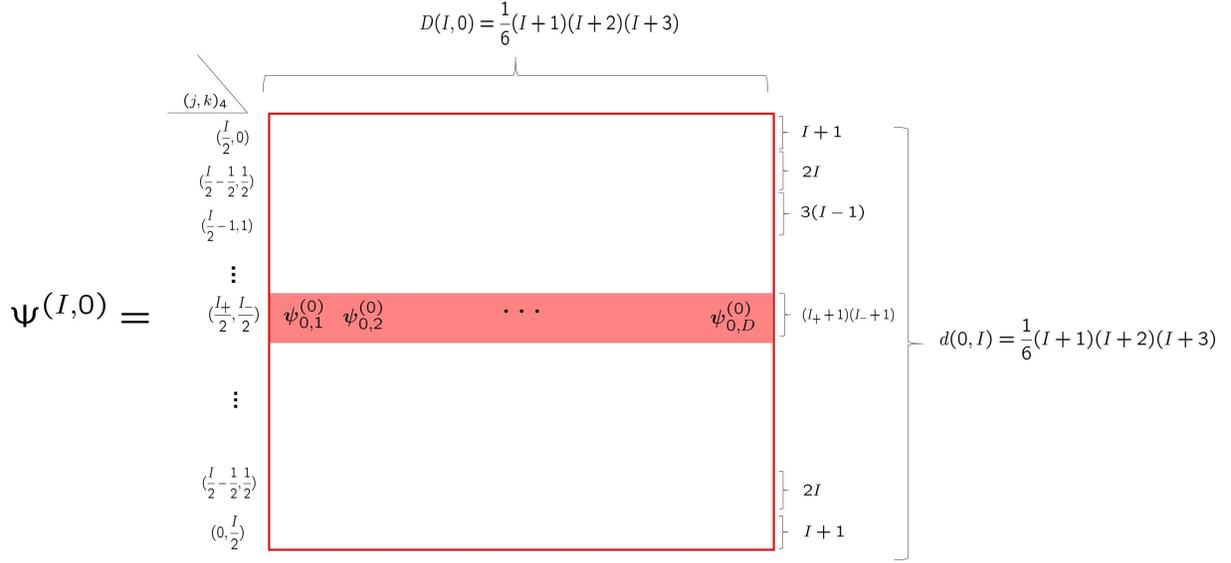}
\caption{ For the $SO(4)$ monopole with $(\frac{I_+}{2}, \frac{I_-}{2})$, the eigenstates of the  LLL  in the $n=0$-sector are realized as the red shaded region of the non-linear realization $\Psi^{(I,0)}$.}
\label{nN0.fig}
\end{figure}

In the Landau problem, the $SO(4)$ monopole indices, $I_+$ and $I_-$, are input parameters, and we need to  specify $p$ and $q$ for the given $I_+$ and $I_-$. The  former two conditions, Eqs.(\ref{ipimfrompqs}) and (\ref{defllfromn}), uniquely specify the $SO(5)$  indices as 
\be
p=N+I-n, ~~~q=N+n,  \label{identifso5}
\ee
where 
\be
I\equiv I_++I_-.
\ee
Since $p\ge q$, Eq.(\ref{identifso5}) implies that $n$ has an upper limit  
and  the range of $n$ may be given by 
\be
n=0, 1, 2, ~\cdots, ~\text{Min}(I_+, I_-). 
\ee
We give a precise  prescription for deriving the $N$th Landau level eigenstates in the $n$-sector and derive several eigenstates. 
We first need to derive the $SO(5)$ matrix generators, $\Sigma_{ab}^{(p,q)_5}$, with  $(p,q)_5=(N+I-n, N+n)_5$. 
That is doable by taking the matrix elements of the $SO(5)$ angular momentum operators with 
 Yang's monopole harmonics  as discussed in Sec.\ref{subsec:arbitso5mat}. Next from the matrix generators,  we construct the non-linear realization using the formula, 
\be
\Psi(\Omega_4) = \exp(i\xi\sum_{m=1}^4 y_m\Sigma_{m5}^{(p,q)_5})\biggl|_{p=N+I-n, ~q=N+n}. \label{genenonline}
\ee
Finally, as indicated in Fig.\ref{geneig.fig}, we  extract an appropriate block matrix from the $n$-sector of $\Psi$.  
The components 
$\bs{\psi}_{N, \alpha}^{(n)}$ denote the $N$th Landau level eigenstates in the $n$-sector, which are normalized 
as\footnote{The connection of $\bs{\psi}_{N,\alpha}^{(n)}$ yields the $SO(4)$ monopole gauge field  (\ref{so4monoinput}), 
\be
A=-i\sum_{\alpha=1}^{D(p,q)} {\bs{\psi}_{N,\alpha}^{(n)}}{d \bs{\psi}_{N,\alpha}^{(n)}}^{\dagger}
= -\frac{1}{1+x_5}\sigma_{mn}^{(\frac{I_+}{2}, \frac{I_-}{2})_4}x_ndx_m .  \label{so4monogau}
\ee
Note that the $A$ in (\ref{so4monogau}) does not depend on either  $N$ or  $n$.}  
\be
\sqrt{\frac{D(p,q)}{(I_++1)(I_-+1) ~{A}(S^4) }}~\bs{\psi}_{N, \alpha}^{(n)}(\Omega_4), 
\ee
with $A(S^4)={8\pi^2}/{3}$. Especially for the LLL $(N=0)$ in the $n=0$-sector,  the eigenstates are given by the red shaded region in  Fig.\ref{nN0.fig}.  

\begin{figure}[tbph]
\center
\includegraphics*[width=160mm]{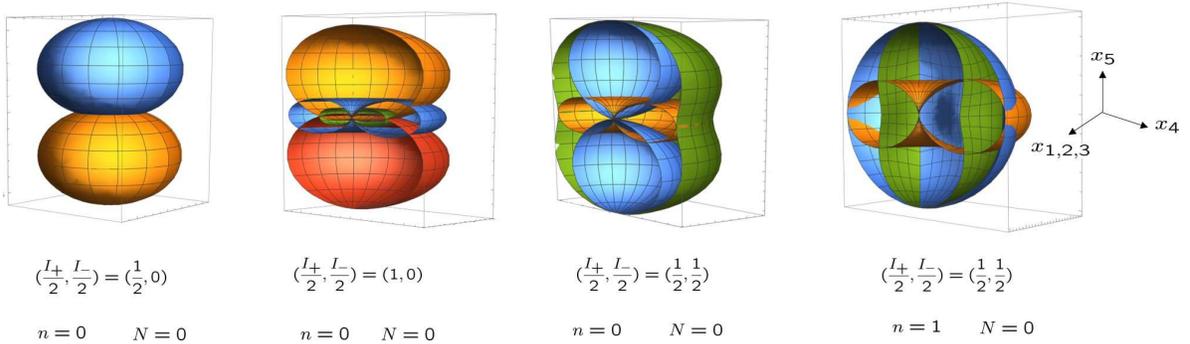}
\caption{
Probability densities of the $SO(5)$ monopole harmonics.  The different colored probability densities correspond to  distinct $SO(5)$ monopole harmonics. For $I_+ \neq I_-$ (the left two) each of the  probability distributions is asymmetric with respect to $x_5=0$ in general, while for $I_+= I_-$ (the right two) each  probability distribution is symmetric with respect to $x_5=0$. 
}
\label{probden.fig}
\end{figure}

Mathematical software is highly efficient in  practically deriving    the non-linear realization. Computation time will be significantly reduced  using the Euler decomposition form of (\ref{genenonline}):  
\be
\Psi(\Omega_4) 
 =
H(\Omega_3)^{\dagger}~e^{i\xi\Sigma_{45}}~ H(\Omega_3)  
\label{pqdfuncso5}
\ee
where 
\be
H(\Omega_3) =e^{-i\chi\Sigma_{34}} e^{i\theta\Sigma_{31}} e^{i\phi\Sigma_{12}}.  
\ee
Following the above prescription,  
 we have derived  the $SO(5)$ monopole harmonics in several $SO(4)$ monopole backgrounds (see Appendix \ref{appendix:trsmonoharm} also), and  
  their probability densities are depicted  in Fig.\ref{probden.fig}.  

\subsection{$SO(5)$ Landau levels}\label{subsec:so5levels}

\begin{figure}[tbph]
\center 
\includegraphics*[width=160mm]{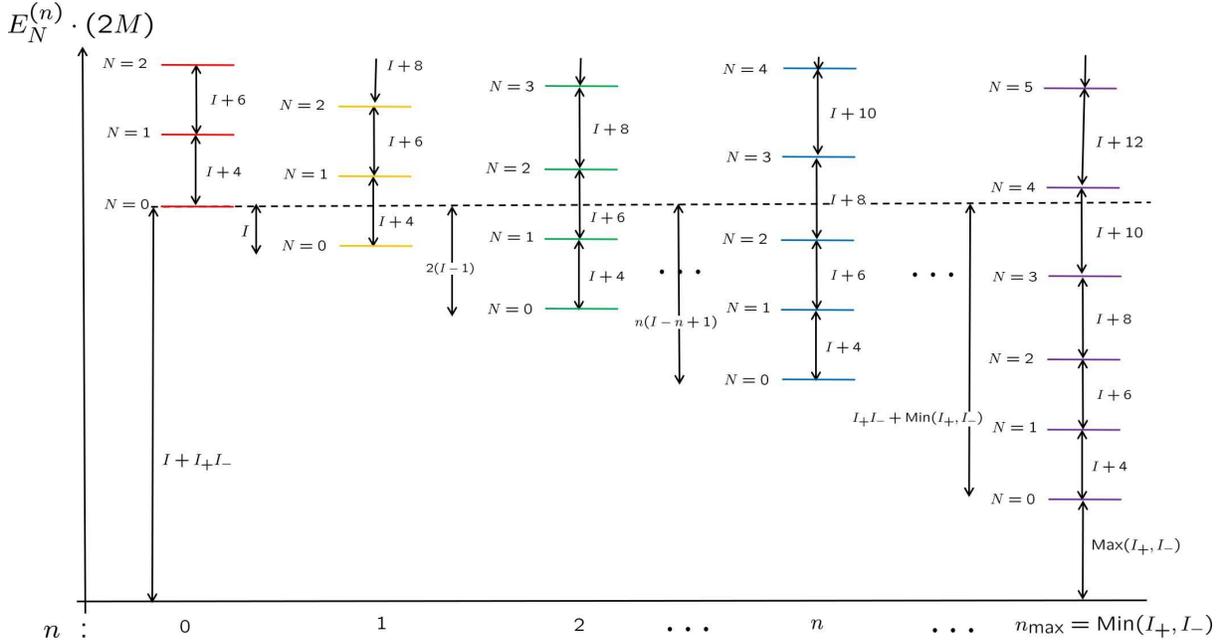}
\caption{For the $SO(4)$ monopole with $(\frac{I_+}{2}, \frac{I_-}{2})_4$, there are $\text{Min}(I_+, I_-)+1$ sectors, each of which exhibits the Landau levels.  
}
\label{genll.fig}
\end{figure}

With (\ref{identifso5}), we may derive the energy levels of 
 (\ref{so5so4monolls}) as 
\begin{align}
E_{N}^{(n)}&=\frac{1}{2M}\biggl(\lambda(p,q)|_{(p,q) =(N+I-n, N+n)} - \frac{1}{2}({I_+}({I_+}+2)+{I_-}({I_-}+2))\biggr) \nn\\
& =\frac{1}{2M}(N(N+3) +I(N-n) +n(n-1))+\frac{1}{2M}(I+I_+I_-), \label{energyso5nll}
\end{align}
where 
\be 
N=0,1,2,\cdots  ~~~\text{and}~~~n=0,1,2,\cdots, \text{Min}(I_+, I_-). 
\ee
Since all possible $(p, q)_5$ are exploited  by changing $N$ and $n$ in (\ref{identifso5}) for a given $I$, Eq.(\ref{energyso5nll}) exhausts all  energy levels of the $SO(5)$ Landau Hamiltonian. 
 Figure \ref{genll.fig} schematically depicts the energy levels  of (\ref{energyso5nll}). 
The corresponding degeneracy (\ref{so5degelandau}) is also derived as  
\be
D_N^{(n)}(I) = \frac{1}{6}(N+n+1)(I-2n+1)(I+N+2-n)(I+2N+3). \label{degenll}
\ee

We here mentioned  specific features of the energy levels.    
The original Landau levels in the $SU(2)$ monopole background  correspond to  the $n=0$ sector of the the preset energy levels.  Indeed, 
for $(I_+, I_-)=(0, I)$ and $n=0$, the above formulas exactly reproduce  the results of Sec.\ref{subsec:so5ori}.     
The Landau level spacing and the degeneracy  depend only on the sum $I\equiv I_+ +I_-$ rather than both $I_+$ and $I_-$. Furthermore, 
the Landau level spacing does not depend on the sector index $n$ and is common in all of the sectors: 
\be
E_{N+1}^{(n)} -E_N^{(n)} =\frac{1}{2M} (2N+I+4). \label{llspacing}
\ee
The Landau level energy monotonically lowers as $n$ increases,  
\be
E_{N}^{(n+1)} -E_N^{(n)} =-\frac{1}{2M} (I-2n) \le 0, 
\ee
and the minimum  energy level is realized at  the LLL of the $n_{\text{max}}=\text{Min}(I_+, I_-)$ sector, 
\be
E_{N=0}^{(n=n_{\text{max}})} =\frac{1}{2M}~\text{Max}(I_+, I_-).  \label{minieneso5}
\ee
Recovering the radius $R$ of the $S^4$ in (\ref{energyso5nll}), we take  the thermodynamic limit, $I, R ~\rightarrow~\infty$  with $I/{R^2}$ being fixed. From (\ref{llspacing}), we see  that every Landau level spacing in all  sectors   becomes identical,    
\be
E_{N+1}^{(n)}-E_{N}^{(n)} ~~\rightarrow~~\omega \equiv \frac{I}{2MR^2} , \label{thermodynali}
\ee
which  is the usual Landau level spacing on a (4D) plane.

\section{Non-commutative geometry and many-body wavefunction}\label{sec:noncommgeo}

Here, we investigate matrix geometries in the Landau levels by applying the Landau level projection \cite{Hasebe-2020-1,  Hasebe-2015, Hasebe-2018}. With the $N$th Landau level eigenstates in the $n$-sector, we take matrix elements of the $S^4$-coordinates: 
\be
(X_a)_{\alpha\beta} =\int d\Omega_4 ~{\bs{\psi}_{N, \alpha}^{(n)}}^{\dagger}~x_a ~ {\bs{\psi}_{N, \beta}^{(n)}}. ~~~~~~~
(\alpha, \beta= 1, 2, \cdots , D_N^{(n)})  \label{matlargx}
\ee
We introduce the $(I_++1)(I_-+1)\times D_N^{(n)}(I)$ matrix that represents the  blue shaded region in Fig.\ref{geneig.fig}: 
\be
\Psi^{(n)}_N \equiv 
 \begin{pmatrix} 
\bs{\psi}_{N, 1}^{(n)}  & \bs{\psi}_{N, 2}^{(n)}  & \cdots & \bs{\psi}_{N, D_N^{(n)}}^{(n)} 
\end{pmatrix}, \label{shadepsima}
\ee
which satisfies 
\be
\Psi^{(n)}_N {\Psi^{(n)}_N }^{\dagger} =\bs{1}_{(I_+ +1)(I_-+1)}. \label{proppsimatrecta}
\ee
Using a $D_N^{(n)}(I)\times D_{N}^{(n)}(I)$ projection matrix $P_N^{(n)}$ made of (\ref{shadepsima})\footnote{
The  $P_N^{(n)}$ holds two eigenvalues, 1 and 0, with degeneracies,  $(I_++1)(I_-+1)$ and $D_N^{(n)}(I)-(I_++1)(I_-+1)$.   
}
\be
P_{N}^{(n)} \equiv {\Psi^{(n)}_N}^{\dagger} \Psi^{(n)}_N ~~~~~~~({P_N^{(n)}}^2=P_N^{(n)})   \label{projemat}
\ee
we can concisely represent the matrix coordinates (\ref{matlargx})  as 
\be
X_a =\int d\Omega_4 ~x_a~P_{N}^{(n)},  \label{matrixxint}
\ee
which obviously signifies the 
projection of the $S^4$-coordinates  to the level.

\subsection{The non-chiral LLL in the $n=I/2$ sector}

Let us first consider the non-chiral LLL eigenstates of the  $n_{\text{max}}=I/2$-sector in the $SO(4)$ monopole background with $(I_+, I_-) =({I}/{2}, {I}/{2})$ $(I: \text{even})$. 
While the second Chern number vanishes,  the zero-point energy $I/(4M)$ is finite and   
 the LLL degeneracy 
 is large as given by $D_{N=0}^{(n=\frac{I}{2})}(I)=\frac{1}{24}(I+2)(I+3)(I+4)$. Therefore, even though the second Chern number is zero, the non-chiral $SO(4)$ monopole  system is not quite the same as  a simple free system  without  $SO(4)$ monopole.  The LLL eigenstates  constitute 
\be
(p,q)_5 =(\frac{I}{2},\frac{I}{2})_5  ,  \label{pqi2}
\ee
and $x_a$ are  
\be
(p,q)_5 =(1,1)_5, 
\ee
so the $SO(5)$ decomposition rule for $x_a\bs{\psi}$  in (\ref{matlargx}) signifies\footnote{See \cite{Hasebe-2020-1} and references therein.}   
\be
(1,1)_5 \otimes (\frac{I}{2},\frac{I}{2})_5 = (\frac{I}{2}+1,\frac{I}{2}+1)_5 \oplus (\frac{I}{2}+1,\frac{I}{2}-1)_5 \oplus (\frac{I}{2}-1,\frac{I}{2}-1)_5. \label{11i2}
\ee
The LLL irreducible representation (\ref{pqi2}) does not exist on the right-hand side of (\ref{11i2}), and then  
\be
X_a =0. \label{zeroxa}
\ee
An intuitive explanation for this result is as follows. For non-chiral cases   (see the right two of Fig.\ref{probden.fig}),  the ``center'' of every probability distribution  is  at the origin,  and hence the expectation values of the coordinates for such states are expected to be  zeros  as in  the case of the spherical harmonics. 
Careful readers may derive the projection matrix (\ref{projemat}) and explicitly check  (\ref{zeroxa}) by performing the integration (\ref{matrixxint}).

For non-chiral LLL eigenstates,  we explicitly computed the Fisher information metric 
\begin{align}
g_{\mu\nu}&=\text{tr}\biggl(\sum_{\alpha=1}^D (\partial_{\mu}\bs{\psi}_{\alpha}  \partial_{\nu}\bs{\psi}_{\alpha}^{\dagger}  + \partial_{\nu}\bs{\psi}_{\alpha}  \partial_{\mu}\bs{\psi}_{\alpha}^{\dagger})-  2  \sum_{\alpha, \beta=1}^D \partial_{\mu}\bs{\psi}_{\alpha} \bs{\psi}_{\alpha}^{\dagger}~ \bs{\psi}_{\beta}\partial_{\nu}\bs{\psi}_{\beta}^{\dagger}\biggr) \nn\\ 
&=\text{tr}\biggl(\partial_{\mu}\Psi_N^{(n)}  \partial_{\nu}{\Psi_N^{(n)}}^{\dagger}  + \partial_{\nu}\Psi_N^{(n)}  \partial_{\mu}{\Psi_{N}^{(n)}}^{\dagger}-  2   \partial_{\mu}\Psi_N^{(n)}  {\Psi_{N}^{(n)}}^{\dagger}~ \Psi_{N}^{(n)}\partial_{\nu}{\Psi_{N}^{(n)}}^{\dagger}\biggr)   \label{fishermetge}
 ~~~~(\mu,\nu=\xi, \chi, \theta, \phi)
\end{align}
to have 
\be
g_{\mu\nu} ~\propto ~\text{diag}(1,~ \sin^2\xi,~\sin^2\xi~\sin^2\chi, ~\sin^2\xi~ \sin^2\chi~\sin^2\theta), \label{mets4angl}
\ee
which is the polar coordinate metric on $S^4$. This is the same result as the $SU(2)$ monopole case \cite{Ishiki-Matsumoto-Muraki-2018} whose fuzzy geometry is the fuzzy four-sphere.  
The Fisher metric reflects the information of the manifold on which the wavefunctions are defined, while the matrix geometry reflects the shapes of the wavefunctions also.

\subsection{The LLL in the $n=0$-sector}

Next, we proceed to   the matrix geometry of the LLL in the $n=0$-sector with  degeneracy  
\be
D_{N=0}^{(n=0)}(I) =\frac{1}{6}(I+1)(I+2)(I+3) ~~~(I\equiv I_+ +I_-) \label{fullysymdimso5}
\ee
and $X_a$ (\ref{matlargx}) are represented by  $D_{N=0}^{(n=0)}(I) \times D_{N=0}^{(n=0)}(I)$ matrices.  

For the $SU(2)$ monopole $(I_+=I,~ I_-=0)$, the previous studies \cite{Hasebe-2020-1, Ishiki-Matsumoto-Muraki-2018}  showed the emergent matrix geometry is  the fuzzy four-sphere:  
\be
X_a =\frac{1}{I+4}~\Gamma_a^{(I)}, \label{su2matgeo}
\ee
where $\Gamma_a^{(I)}$ are fully symmetric  tensor products of $I$ $SO(5)$ gamma matrices.\footnote{In particular, 
\be
\Gamma_m^{(I=1)}=\begin{pmatrix}
0 & \bar{q}_m \\
q_m & 0 
\end{pmatrix}=\gamma_m, ~~~\Gamma_5^{(I=1)}=\begin{pmatrix}
1_2 & 0 \\
0 & -1_2 
\end{pmatrix}=-\gamma_5. 
\ee
}   
The basic properties of $\Gamma_a^{(I)}$ are  given by 
\be
[\Gamma_a, \Gamma_b, \Gamma_c, \Gamma_d] =8(I+2) \epsilon_{abcde}\Gamma_e,  ~~~~~~\Gamma_a^{(I)}\Gamma_a^{(I)}=I(I+4)~\bs{1}_{\bs{1}_{D_{N=0}^{(n=0)}(I)}},  
\ee 
where the four bracket $[~,~,~,~]$ denotes the fully antisymmetric combinations of the four quantities inside the bracket: 
\be
[\Gamma_a, \Gamma_b, \Gamma_c, \Gamma_d] \equiv \sum_{\sigma }\text{sgn}(\sigma) ~\Gamma_{\sigma(a)}\Gamma_{\sigma(b)}\Gamma_{\sigma(c)}\Gamma_{\sigma(d)}. 
\ee

For the $SO(4)$ monopole background with index $(\frac{I_+}{2}, \frac{I_-}{2})_4$, we explicitly evaluate  $X_a$ (\ref{matrixxint}) using several low dimensional representations. 
From the obtained results, we deduce that the matrix geometry in the $SO(4)$ monopole background becomes  
\be
X_a = \frac{I_+-I_-}{I~(I+4)}~\Gamma_a^{(I)}.  \label{genematgeom}
\ee
This naturally generalizes the original result (\ref{su2matgeo}). Notice that the matrix size of $X_a$ depends only on the sum of  the $SO(4)$ bi-spin indices while the overall coefficient  depends on the  difference of the $SO(4)$ bi-spin indices.   
The matrix coordinates (\ref{genematgeom}) satisfy the quantum Nambu geometry of the fuzzy four-sphere, 
\be
[X_a, X_b, X_c, X_d] =(I+2)\biggl( \frac{2(I_+-I_-)}{I(I+4)} \biggr)^3 \epsilon_{abcde}X_e, 
\ee
and the radius is  
\be
X_aX_a =\frac{(I_+-I_-)^2}{I~(I+4)}~ \bs{1}_{D_{N=0}^{(n=0)}(I)}. \label{xaxaradius}
\ee
Equation (\ref{xaxaradius}) implies that  the monopole and anti-monopole  oppositely contribute to the radius of the fuzzy four-sphere, and notably at the non-chiral case $I_+ =I_-$, the radius  apparently vanishes. 
The Fisher metric  is again given by the classical four-sphere metric  (\ref{mets4angl}).

\subsection{4D quantum Hall wavefunction}

 The non-commutative geometry is the underlying geometry of the quantum Hall effect and governs the LLL physics \cite{Girvin-Jach-1984, Girvin-MacDonald-Platzman-1986, Ezawa-Tsitsishvili-Hasebe-2003}.
As the LLL  geometry    in the $n=0$-sector  is given by the fuzzy four-sphere geometry same as the original 4D quantum Hall effect, a Laughlin-like many-body wavefunction is 
 expected to be realized in the present system. Recall that in the original 4D quantum Hall effect \cite{Zhang-Hu-2001}, the many-body wavefunction is  constructed as the $m$th power of the Slater determinant,  
\be 
 \Psi^{(m)}(x_1,x_2,\cdots,x_{D}) =\Psi_{\text{Slat}}(x_1,x_2,\cdots, x_{i_{D}})^m,  
\ee 
where 
\begin{equation}
\Psi_{\text{Slat}}(x_1,x_2,\cdots, x_{D}) =\epsilon_{i_1 i_2 \cdots i_{D}} ~\bs{\psi}_{1}(x_{i_1}) ~\bs{\psi}_{2}(x_{i_2})~ \cdots  ~\bs{\psi}_{D} (x_{i_{D}}).  \label{slatemany}
\end{equation}
The symbol $m$ is taken to be an odd integer due to the Fermi statistics. The right-hand side of (\ref{slatemany}) is  the tensor products  
of Yang's LLL monopole harmonics  with  degeneracy $D=\frac{1}{6}(I+1)(I+2)(I+3)$. 
Since Yang's LLL monopole harmonics are given by the symmetric products  of the $SO(5)$ fundamental spinors, it is legitimate to adopt $\Psi^{(m)}$ as a Laughlin-like many-body function \cite{Zhang-Hu-2001}. 
We see that the power of each one-particle state  is equally given by $mI$, which implies the corresponding $SU(2)$ monopole index to be  $m\frac{I}{2}$.

 In the same spirit, we construct a Laughlin-like many-body wavefunction for the LLL of the $n=0$-sector in the $SO(4)$ monopole background with indices,  
\be
(m\frac{I_+}{2}, ~m\frac{I_-}{2})_4.  \label{mso4indices}
\ee
The filling factor  is given by 
\be
\nu =\frac{D_{N=0}^{(n=0)}(I_++I_-)}{D_{N=0}^{(n=0)}(mI_+ +mI_-)} ~~\overset{I_+ +I_-~~\rightarrow~~\infty} {\longrightarrow}~~\frac{1}{m^3}. \label{fillinggene}
\ee
It is straightforward to derive the Slater determinant wavefunction at  filling $\nu=1$ using the LLL monopole harmonics in the $n=0$-sector. 
The obtained Slater determinant is a singlet under the $SO(5)$ rotations 
and represents a uniformly distributed non-interacting many-body  state on a four-sphere. 
However, in the construction  of the Laughlin wavefunction, the situation is rather involved; powers of the Slater determinant are $\it{not}$ generally confined in the LLL.   
This is because  the LLL one-particle states in the $SO(4)$ monopole background  are not simply given by homogeneous polynomials unlike the original $SU(2)$ case.    
Therefore, we have to implement the  projection to the LLL, 
\be
\Psi^{(m)}(x_1,x_2,\cdots,x_{D})=\text{P}_{\text{LLL}}\Psi_{\text{Slat}}(x_1,x_2,\cdots, x_{i_{D}})^m, \label{trulllfunc}
\ee
where $\text{P}_{\text{LLL}}$ denotes the projection operator  constructed by 
\be
\text{P}_{\text{LLL}}=\sum_{\text{singlet}}|\text{singlet}\rangle \langle \text{singlet}|.  
\ee
The states $|\text{singlet}\rangle$  signify the $SO(5)$ singlets 
 made of  the $D_{N=0}^{(n=0)}(I)$ tensor products of  the LLL monopole harmonics in the $n=0$-sector with the $SO(4)$ background of indices (\ref{mso4indices}). Applying the projection operator,  we extract the LLL components of the $m$th power of the Slater determinant  not ruining the $SO(5)$ symmetry.  In this way,   we can construct a Laughlin-like many-body groundstate  at filling (\ref{fillinggene}).

\section{Relativistic $SO(5)$ Landau model}\label{sec:relativisticlandau}

We explore the relativistic version of the $SO(5)$ Landau model for a spinor particle and demonstrate 
the Atiyah-Singer index theorem for the $SO(4)$ monopole gauge field. 

\subsection{Synthetic gauge field and the relativistic Landau levels}

With   
\be
\omega_{mn} =\frac{1}{1+x_5}(x_m dx_n -x_n dx_m), \label{coordomega}
\ee  
the spin connection of $S^4$ is given by\footnote{
The matrices of (\ref{spinmatconn}) are  
\be
\sigma_{mn}^{(\frac{1}{2}, 0)_4}=\frac{1}{2}\eta_{mn}^i\sigma_i,~~~\sigma_{mn}^{(0,\frac{1}{2})_4}=\frac{1}{2}\bar{\eta}_{mn}^i\sigma_i. \label{omegagaufield1}
\ee
} 
\be
\omega  = \frac{1}{2} \omega_{mn}(\sigma_{mn}^{(\frac{1}{2}, 0)_4}\oplus {\sigma}_{mn}^{(0, \frac{1}{2})_4}),\label{spinmatconn}
\ee
and the  $SO(4)$ monopole gauge field (\ref{so4monoinput}) is 
\be
A=\frac{1}{2}\omega_{mn}\sigma_{mn}^{(\frac{I_+}{2},\frac{I_-}{2})_4}. \label{gaufield1}
\ee 
The relativistic $SO(5)$ Landau model describes  a spinor particle on $S^4$, which interacts with the $SO(4)$ gauge field and  the spin connection as well, and so their synthetic connection is the concern    
\be 
\mathcal{A} 
 \equiv \omega\otimes \bs{1}_{(I_++1)(I_-+1)} +1_4\otimes A. 
\ee
The Dirac-Landau operator on $S^4$ is constructed as 
\be
-i\fsl{\mathcal{D}}
=-i\gamma^m e_m^{~~\mu}(\partial_{\mu}  +i\mathcal{A}_{\mu}), 
\ee
where $\mu$ denote the local coordinates on $S^4$, such as $\xi, \chi, \theta, \phi$.  

Since the coordinate-dependent parts of $\omega$ and $A$   are identical (\ref{coordomega}),\footnote{Recall that we have chosen the gauge group as the holonomy group of $S^4$.} the synthetic gauge field is simply obtained  by taking the tensor product of   the $SO(4)$ matrices of (\ref{spinmatconn}) and (\ref{gaufield1}). 
According to the $SO(4)$ decomposition rule\footnote{Since  $SO(4)\simeq   SU(2)\otimes SU(2)$, we can apply the $SU(2)$  decomposition rule to each of the  $SU(2)$s: 
\be
(j, k)_4\otimes (j', k')_4 =\bigoplus_{J=|j-j'|}^{j+j'} \bigoplus_{K=|k-k'|}^{k+k'} ~ (J, K)_4,
\ee
or 
\be
\sigma_{mn}^{(j,k)_4} \otimes 1_{(2j'+1)(2k'+1)}
+ 1_{(2j+1)(2k+1)}
\otimes \sigma_{mn}^{(j',k')_4} 
=\sigma_{mn}^{(j\otimes j',k\otimes k')_4}
=
 \bigoplus_{J=|j-j'|}^{j+j'} \bigoplus_{K=|k-k'|}^{k+k'} ~\sigma^{(J,K)_4}_{mn}. 
\ee
}  
\be 
((\frac{1}{2},  0)_4 \oplus (0, \frac{1}{2})_4) \otimes (\frac{I_+}{2}, \frac{I_-}{2})_4 \nn\\
= (\frac{I_+ }{2} + \frac{1}{2} , \frac{I_-}{2})_4 \oplus (\frac{I_+ }{2} , \frac{I_-}{2}  + \frac{1}{2})_4 \oplus (\frac{I_+ }{2}-\frac{1}{2} , \frac{I_-}{2})_4 \oplus (\frac{I_+ }{2} , \frac{I_-}{2}-\frac{1}{2})_4 , 
\label{so4decomprule}
\ee 
we see that  the synthetic connection consists of the four sectors: 
\be 
\mathcal{A}^{(\frac{\mathcal{I}_+}{2}, \frac{\mathcal{I}_-}{2})_4} ~=
~A^{(\frac{I_+}{2} +\frac{1}{2} , \frac{I_-}{2})_4} \oplus A^{(\frac{I_+}{2} , \frac{I_-}{2}  +\frac{1}{2})_4} \oplus A^{(\frac{I_+}{2}-\frac{1}{2} , \frac{I_-}{2})_4} \oplus A^{(\frac{I_+}{2}  , \frac{I_-}{2}-\frac{1}{2} )_4}.  
\label{so4decomprulegauge}
\ee 
A standard way for  deriving the spectra of the Dirac-Landau operator is to take its square  and make use  of the results of the corresponding non-relativistic Landau problem.  
The formula is given by \cite{Dolan-2003, Hasebe-2014-1} 
\be
(-i\fsl{\mathcal{D}})^2 =\sum_{a<b}{\mathcal{L}_{ab}}^2 -\sum_{a<b}{F_{ab}}^2 +\frac{1}{4}{R}_{S^4}. \label{squdilanequal}
\ee 
The symbol $R_{S^4}=6$ is the scalar curvature of $S^4$, and $\mathcal{L}_{ab}$ denote the angular momentum operators with the synthetic gauge field 
\be
\mathcal{L}_{ab}=-ix_{a}(\partial_b +i\mathcal{A}_b) +ix_b(\partial_a +i\mathcal{A}_b) +r^2\mathcal{F}_{ab},  
\ee
where $\mathcal{F}_{ab} =\partial_a \mathcal{A}_b-\partial_{b}\mathcal{A}_a +i[\mathcal{A}_a, \mathcal{A}_b]$. The operators $\mathcal{L}_{ab}$ are just the familiar $SO(5)$ angular momentum operators with the $SO(4)$ monopole gauge field of the indices $(\frac{\mathcal{I}_+}{2}, \frac{\mathcal{I}_-}{2})$. We apply the results of Sec.\ref{subsec:so5levels}  to derive the spectra  
\be 
(-i\fsl{\mathcal{D}})^2 
=\lambda(p, q)   -\frac{1}{2}(I_+ (I_+ +2) + I_-(I_-+2)) +\frac{3}{2} ~~\ge 0. 
\ee 
Similar to (\ref{identifso5}),  the $SO(5)$ indices $p$ and $q$ are given by  
\be
p =N+\mathcal{I} -n, ~~~~~q=N+n
\ee
where 
\be
\mathcal{I}\equiv \mathcal{I}_+ + \mathcal{I}_-, ~~~~~~~n=0, 1, 2,   \cdots, ~\text{Min}(\mathcal{I}_+, \mathcal{I}_-).  
\ee 
In the first two cases of (\ref{so4decomprulegauge}), we have $\mathcal{I}=\mathcal{I}_+ + \mathcal{I}_- +1 =I+1$,  
and then 
\be 
-i\fsl{\mathcal{D}} 
=\pm \sqrt{N(N+3) +(I+1)(N-n) +n(n-1) +{2}I +I_+ I_- +4},   \label{biggerener}
\ee 
in which each of the  positive and negative Landau levels holds the same degeneracy 
\be
D(N+I+1-n, N+n) =\frac{1}{6}(N+n+1)(I-2n+2)(I+N+3-n)(I+2N+4). 
\ee
The minimum energy eigenvalue in magnitude is achieved at $N=0$, $n=\text{Min}(\mathcal{I}_+, \mathcal{I}_-)$ to yield  $|-i\fsl{\mathcal{D}}|=\sqrt{2 \text{Max}(I_+, I_-)+4}$, and  the spectra (\ref{biggerener}) do not realize zero-modes.   
Meanwhile in the last two cases of (\ref{so4decomprulegauge}), we have $\mathcal{I}=\mathcal{I}_+ + \mathcal{I}_- -1 =I-1$; 
\be 
-i\fsl{\mathcal{D}} 
=\pm \sqrt{N(N+3) +(I-1)(N-n) +n(n-1) +I_+I_-},  \label{diracopeigseclo}
\ee 
in which  each of the  positive and negative Landau levels of (\ref{diracopeigseclo}) holds the same degeneracy  
\be
D(N+I-1-n, N+n)=\frac{1}{6}(N+n+1)(I-2n)(I+N+1-n)(I+2N+2). \label{dimdegelow}
\ee
For fixed $N$, $n$, $I_+$ and $I_-$, the eigenvalues of  (\ref{diracopeigseclo}) are smaller than those of (\ref{biggerener}) in magnitude and realize zero-modes at  $N=0$, $n=n_{\text{max}}=\text{Min}(\mathcal{I}_+, \mathcal{I}_-)$.

\subsection{Zero-modes  and the Atiyah-Singer index theorem}\label{subsec:asind}

 The Atiyah-Singer index theorem signifies   equality between the zero-mode number and the Chern number.\footnote{
Since the 
 Dirac genus  of sphere is trivial,  we only need to take into account the Chern number in (\ref{aseqtobedet}). 
}    
For the present system, the Atiyah-Singer index theorem may be expressed as
\be
\text{ind}(-i\fsl{\mathcal{D}}) \equiv  \dim \text{Ker}(-i\fsl{\mathcal{D}}_+) -\dim \text{Ker}(-i\fsl{\mathcal{D}}_-) =c_2,  \label{aseqtobedet}
\ee
where $\fsl{\mathcal{D}}_{\pm}$ are defined as  
\be
\fsl{\mathcal{D}}_{\pm } =\frac{1}{2}(1 \pm  \gamma_5)\fsl{\mathcal{D}} 
\ee
and $c_2$ is the second Chern number of the $SO(4)$ monopole (\ref{c2=g}). 
We  evaluate the left-hand side of (\ref{aseqtobedet}) to validate (\ref{aseqtobedet}). 

For $I_+ > I_-$,  the zero-modes are  realized as those of  $-i\fsl{\mathcal{D}}_+$ in  $(\frac{\mathcal{I}_+}{2}, \frac{\mathcal{I}_-}{2}) = (\frac{I_+-1}{2}, \frac{I_-}{2})$ at $N=0$ and $n=\text{Min}(\mathcal{I}_+, \mathcal{I}_-)=I_-$. 
We then find 
$\dim \text{Ker}(-i\fsl{\mathcal{D}}_+) = D(I_+-1, I_-)$ and  $\dim \text{Ker}(-i\fsl{\mathcal{D}}_-) = 0$: 
\be
\text{ind}(-i\fsl{\mathcal{D}})  =D(I_+-1, I_-) =\frac{1}{6}(I_+ +1)(I_-+1)(I_++I_-+2)(I_+-I_-). \label{inddresult}
\ee
Similarly for $I_+ < I_-$, the zero-modes are  realized as those of $-i\fsl{\mathcal{D}}_{-}$ in  $(\frac{\mathcal{I}_+}{2}, \frac{\mathcal{I}_-}{2}) = (\frac{I_+}{2}, \frac{I_- -1}{2})$ at $N=0$ and $n=\text{Min}(\mathcal{I}_+, \mathcal{I}_-)=I_+$. 
We then have 
$\dim \text{Ker}(-i\fsl{D}_+) = 0$ and $\dim \text{Ker}(-i\fsl{D}_-) = D(I_--1, I_+)=-D(I_+-1, I_-)$,  
and so $\text{ind}(-i\fsl{D}) =D(I_+-1, I_-)$,  which yields   
(\ref{inddresult}) again. 
Finally in the case $I_+=I_-=\frac{I}{2}$ $(I=2,4,6,\cdots)$, the LLL of the $n_{\text{max}}=\frac{I}{2}-1$-sector 
(\ref{diracopeigseclo}) does not realize the zero modes ($(-i\fsl{\mathcal{D}}) =\pm 1\neq 0$), $i.e.$, $\dim \text{Ker}(-i\fsl{\mathcal{D}}) = 0$,  
which is also realized at $I_+ =I_-$ in  (\ref{inddresult}).  
After all, for arbitrary $SO(4)$ indices,   Eq.(\ref{inddresult}) generally holds and the most right-hand side is  exactly equal to the second Chern number   (\ref{c2=g}).  This obviously demonstrates the Atiyah-Sinder index theorem. 

\section{Summary and discussions}\label{sec:summary}
  
  In this work, we fully solved the $SO(5)$ Landau problem in the $SO(4)$ monopole background and explored  non-commutative geometry and  4D quantum Hall effect. 
For the $SO(4)$ monopole  with a bi-spin index, $(\frac{I_+}{2}, \frac{I_-}{2})$, 
we demonstrated that the $SO(5)$ Landau model is endowed with $\text{Min}(I_+, I_-)$ sectors,  each of which hosts the Landau levels whose level spacing is determined  by the sum of the $SO(4)$ bi-spins (Fig.\ref{genll.fig}). It was  shown that the $N$th Landau level eigenstates in  the $n$-sector can be obtained as a block matrix of   the non-linear realization (the blue shaded block matrix in Fig.\ref{geneig.fig}) with  
\be
(p, q)_5= (N+I_+ + I_--n, N+n)_5 .
\ee
 The  matrix geometry of the LLL in the $n=0$-sector was identified  as the fuzzy four-sphere whose  radius  is determined by the difference between the $SO(4)$ bi-spin indices, while the  matrix geometry  of the non-chiral  case  is  trivial.  
The classical $S^4$ geometry was recovered as the Fisher information metric in any cases. 
We constructed the Slater determinant from the newly obtained monopole harmonics and derive a Laughlin-like many-body  wavefunction in the $SO(4)$ monopole background  by  applying the LLL projection. 
   We also investigated the $SO(5)$ relativistic Landau model and derived the relativistic spectrum and the degeneracy. The number of the zero-modes exactly coincides with the second Chern number of the $SO(4)$ monopole  as anticipated by the Atiyah-Singer index theorem.   
  
The $SO(4)$ monopole is quite unique for its gauge group being  the only semi-simple group among  the $SO(n)$ groups, which 
endows the present system with a particular multi-sector structure of the  Landau levels. 
It may be interesting to speculate experimental realizations of the present model in real condensed matter systems of  synthetic dimensions. Of particular interest will be the  non-chiral case $I_+ =I_-$,  in which the second Chern number vanishes while  
 the generalized  Euler number does not and 
 its physical implications have not been  understood yet.       
There are many to be clarified in the present model itself, such as edge modes, effective field theory and extended excitations. More explorations  will be beneficial  not only for further understanding of higher D topological phases but also for non-commutative geometry and  string theory.

\section*{Acknowledgements}

 This work was supported by JSPS KAKENHI Grant No.~21K03542.

\appendix

\section{The Pontyagin number and the Euler number}\label{appendix:so4monopole}

On the 4D manifold,  
the (first) Pontryagin number $P_1$ and the Euler number $\chi_4$ are introduced as \cite{Eguchi-Freund-1976}
\begin{subequations}
\begin{align}
&P_1(M^4)= \frac{1}{8\pi^2}\int_{M^4} R^{m_1 m_2} R_{m_1 m_2} =\frac{1}{32\pi^2}\int_{M^4} R^{m_1 m_2} R^{m_3 m_4}\tr(X_{m_1m_2} X_{m_3m_4}),  \\ 
&\chi_4(M^4)=\frac{1}{32\pi^2}\int_{M^4} \epsilon^{m_1m_2m_3m_4}R_{m_1 m_2} R_{m_3 m_4}=\frac{1}{128\pi^2}\int_{M^4} \epsilon^{m_3m_4m_5m_6}R^{m_1 m_2} R_{m_3 m_4}\tr(X_{m_1m_2} X_{m_5m_6}), 
\end{align}\label{defsponteul}
\end{subequations}
where $R^{m_1m_2}$ stand for the curvature two-form of the manifold and $X_{m_1 m_2}$ denote the $SO(4)$ adjoint representation matrices: 
\be
(X_{m_1 m_2})_{m_3 m_4}\equiv -i\delta_{m_1 m_3}\delta_{m_2 m_4} +i\delta_{m_1 m_4}\delta_{m_2 m_3}. \label{defxmn}
\ee
The topological quantities  for the gauge field (\ref{twotopinv}) are generalizations of (\ref{defsponteul}) by replacing the curvature two-form of the adjoint representation matrices with the field strength  of arbitrary representation matrices.

For spheres $S^d$, we have 
\be
R_{mn}=e_m\wedge e_n, 
\ee
and (\ref{defsponteul}) becomes  
\be
P_1(S^4)=0,  ~~~~~~~
\chi_4(S^{4})=2. \label{s4quantitopo}
\ee
Equation (\ref{s4quantitopo}) is realized as a special case of  (\ref{cchimonop}) for the $SO(4)$ vector representation $(\frac{I_+}{2}, \frac{I_-}{2})=(\frac{1}{2}, \frac{1}{2})$. 
This is because  $X_{m_1 m_2}$ (\ref{defxmn})  are unitarily equivalent to $\sigma^{(\frac{1}{2} ,\frac{1}{2})_4}_{m_1 m_2}$ (\ref{so4irrepgenemat}): 
\be
\sigma_{m_1 m_2}^{(\frac{1}{2}, \frac{1}{2})_4}=U^{\dagger}~X_{m_1 m_2} ~U, ~~~~
U=\frac{1}{\sqrt{2}}
\begin{pmatrix}
1 & 0 & 0 & -1 \\
i & 0 & 0 & i \\
0 & -1 & -1 & 0 \\
0 & i & -i & 0 
\end{pmatrix}. 
\ee
Consequently, 
\be
P_1(S^4) =c_2^{(\frac{1}{2}, \frac{1}{2})},~~~~\chi_4(S^4) =\frac{1}{2}\tilde{c}_2^{(\frac{1}{2}, \frac{1}{2})}.
\ee

\section{Non-chiral $SO(5)$  monopole harmonics }\label{appendix:trsmonoharm}

For a better understanding, we derive several  $SO(5)$ monopole harmonics.  
We represent the non-linear realization matrix 
(\ref{exp01psi})  as 
\be
\Psi^{(1,0)_5} 
=\begin{pmatrix}
\bs{\psi}^{[0,\frac{1}{2}]}_1 & \bs{\psi}^{[0,\frac{1}{2}]}_2 & \bs{\psi}^{[0,\frac{1}{2}]}_3 & \bs{\psi}^{[0,\frac{1}{2}]}_4 \\
\bs{\psi}^{[0,-\frac{1}{2}]}_1 & \bs{\psi}^{[0,-\frac{1}{2}]}_2 & \bs{\psi}^{[0,-\frac{1}{2}]}_3 & \bs{\psi}^{[0,-\frac{1}{2}]}_4
\end{pmatrix}.
\ee
The upper column quantities,  $\bs{\psi}_1^{[0,\frac{1}{2}]}, \bs{\psi}_2^{[0,\frac{1}{2}]}, \bs{\psi}_3^{[0,\frac{1}{2}]}, \bs{\psi}_4^{[0,\frac{1}{2}]}$, denote the fourfold degenerate LLL eigenstates in the $SU(2)$ monopole background $(\frac{I_+}{2}, \frac{I_-}{2})=(\frac{1}{2}, 0)$: 
\begin{align}
&\bs{\psi}_1^{[0,\frac{1}{2}]}=\sqrt{\frac{1+x_5}{2}}\begin{pmatrix}
1\\
0
\end{pmatrix}=\begin{pmatrix}
\cos\frac{\xi}{2} \\
0
\end{pmatrix}, ~~\bs{\psi}_2^{[0,\frac{1}{2}]}=\sqrt{\frac{1+x_5}{2}}\begin{pmatrix}
0\\
1
\end{pmatrix}=\begin{pmatrix}
0 \\
\cos\frac{\xi}{2} 
\end{pmatrix}, \nn\\
&\bs{\psi}_3^{[0,\frac{1}{2}]}=\frac{1}{\sqrt{2(1+x_5)}}\begin{pmatrix}
x_4 +ix_3\\
-x_2+ix_1
\end{pmatrix}=\begin{pmatrix}
\sin\frac{\xi}{2} (\cos\chi+i\sin\chi\cos\theta) \\
i\sin\frac{\xi}{2}\sin\chi\sin\theta e^{i\phi}
\end{pmatrix}, \nn\\
&\bs{\psi}_4^{[0,\frac{1}{2}]}=\frac{1}{\sqrt{2(1+x_5)}}\begin{pmatrix}
x_2+ix_1 \\
x_4 -ix_3
\end{pmatrix}=\begin{pmatrix}
i\sin\frac{\xi}{2}\sin\chi\sin\theta e^{-i\phi}\\
\sin\frac{\xi}{2} (\cos\chi-i\sin\chi\cos\theta)
\end{pmatrix}, \label{uppertwocomp}
\end{align}
while the lower column quantities, $\bs{\psi}^{[0,-\frac{1}{2}]}_1, \bs{\psi}^{[0,-\frac{1}{2}]}_2, \bs{\psi}^{[0,-\frac{1}{2}]}_3, \bs{\psi}^{[0,-\frac{1}{2}]}_4$, represent 
the fourfold degenerate LLL eigenstates in the $SU(2)$ anti-monopole background  $(\frac{I_+}{2}, \frac{I_-}{2})=(0, \frac{1}{2})$:
\begin{align}
&\bs{\psi}^{[0,-\frac{1}{2}]}_1=\frac{1}{\sqrt{2(1+x_5)}}\begin{pmatrix}
-x_4 +ix_3\\
-x_2+ix_1
\end{pmatrix}=\begin{pmatrix}
-\sin\frac{\xi}{2} (\cos\chi-i\sin\chi\cos\theta) \\
i\sin\frac{\xi}{2}\sin\chi\sin\theta e^{i\phi}
\end{pmatrix} ,\nn\\
&\bs{\psi}^{[0,-\frac{1}{2}]}_2={\frac{1}{\sqrt{2(1+x_5)}}}\begin{pmatrix}
x_2+ix_1 \\
-x_4 -ix_3
\end{pmatrix}=\begin{pmatrix}
i\sin\frac{\xi}{2}\sin\chi\sin\theta e^{-i\phi}\\
-\sin\frac{\xi}{2} (\cos\chi+i\sin\chi\cos\theta)
\end{pmatrix},  \nn\\
&\bs{\psi}^{[0,-\frac{1}{2}]}_3=\sqrt{\frac{1+x_5}{2}}\begin{pmatrix}
1\\
0
\end{pmatrix}=\begin{pmatrix}
\cos\frac{\xi}{2} \\
0
\end{pmatrix}, ~~\bs{\psi}^{[0,-\frac{1}{2}]}_4=\sqrt{\frac{1+x_5}{2}}\begin{pmatrix}
0\\
1
\end{pmatrix}=\begin{pmatrix}
0 \\
\cos\frac{\xi}{2} 
\end{pmatrix}. \label{lowertwocomp}
\end{align}
Following the prescription in the main text,   we can derive the tenfold degenerate LLL $SO(5)$ eigenstates in the $n=0$-sector of the $SO(4)$ background  $(\frac{I_+}{2}, \frac{I_-}{2})_4=(\frac{1}{2}, \frac{1}{2})_4$. From the non-linear realization matrix of $(p,q)_5=(2,0)_5$, we have    
\footnotesize  
\begin{align}
&\bs{\psi}^{[0,0]}_{(1,1)} =\frac{1}{\sqrt{2}}\begin{pmatrix}
-\sin\xi (\cos\chi -i\sin\chi\cos\theta) \\
i\sin\xi\sin\chi\sin\theta e^{i\phi} \\
0 \\
0 
\end{pmatrix},~\bs{\psi}_{(1,2)}^{[0,0]} =\frac{1}{{2}}\begin{pmatrix}
i\sin\xi\sin\chi\sin\theta e^{-i\phi} \\
- \sin\xi (\cos\chi+i\sin\chi\cos\theta)\\
 -\sin\xi (\cos\chi-i\sin\chi\cos\theta) \\
i\sin\xi\sin\chi\sin\theta e^{i\phi}
\end{pmatrix},~\bs{\psi}_{(3,3)}^{[0,0]} =\frac{1}{\sqrt{2}}\begin{pmatrix}
 0 \\
0 \\
i\sin\xi\sin\chi\sin\theta e^{-i\phi}  \\
 -\sin\xi (\cos\chi +i\sin\chi\cos\theta) 
\end{pmatrix},\nn\\
&\bs{\psi}_{(1,3)}^{[0,0]} =
\begin{pmatrix}
\cos^2\frac{\xi}{2}-\sin^2\frac{\xi}{2}(\cos^2\chi +\sin^2\chi \cos^2\theta) \\
i\sin^2\frac{\xi}{2}\sin\chi\sin\theta(\cos\chi+i\sin\chi\cos\theta)e^{i\phi} \\
-i\sin^2\frac{\xi}{2}\sin\chi\sin\theta(\cos\chi-i\sin\chi\cos\theta)e^{i\phi}\\
-\sin^2\frac{\xi}{2}\sin^2\chi\sin^2\theta e^{2i\phi}
\end{pmatrix},~~\bs{\psi}_{(1,4)}^{[0,0]}=
\begin{pmatrix}
-i\sin^2\frac{\xi}{2}\sin\chi\sin\theta(\cos\chi-i\sin\chi\cos\theta)e^{-i\phi}\\
\cos^2\frac{\xi}{2}-\sin^2\frac{\xi}{2}\sin^2\chi \sin^2\theta \\
-\sin^2\frac{\xi}{2}(\cos\chi-i\sin\chi\cos\theta)^2 \\
i\sin^2\frac{\xi}{2}\sin\chi\sin\theta  (\cos\chi-i\sin\chi\cos\theta )e^{i\phi}
\end{pmatrix},\nn\\
&\bs{\psi}_{(2,3)}^{[0,0]} = 
\begin{pmatrix}
i\sin^2\frac{\xi}{2} \sin\chi\sin\theta (\cos\chi+ i\sin\chi\cos\theta)e^{-i\phi} \\
-\sin^2\frac{\xi}{2} (\cos\chi+i\sin\chi\cos\theta)^2 \\
\cos^2\frac{\xi}{2}-\sin^2\frac{\xi}{2}\sin^2\chi \sin^2\theta \\
-i\sin^2\frac{\xi}{2}(\cos\chi+i\sin\chi\cos\theta)\sin\chi\sin\theta 
\end{pmatrix}, ~~\bs{\psi}_{(2,4)}^{[0,0]}=
\begin{pmatrix}
-\sin^2\frac{\xi}{2}\sin^2\chi \sin^2\theta e^{-2i\phi} \\
-i\sin^2\frac{\xi}{2} \sin\chi\sin\theta (\cos\chi+i\sin\chi\cos\theta) e^{-i\phi} \\
i\sin^2\frac{\xi}{2}\sin\chi\sin\theta (\cos\chi-i\sin\chi\cos\theta) e^{-i\phi} \\
\cos^2\frac{\xi}{2}-\sin^2\frac{\xi}{2}(\cos^2\chi +\sin^2\chi \cos^2\theta)
\end{pmatrix}, \nn\\
&\bs{\psi}_{(3,3)}^{[0,0]}=\frac{1}{\sqrt{2}}
\begin{pmatrix} 
\sin\xi (\cos\chi+i\sin\chi\cos\theta) \\
0 \\
i\sin\xi\sin\chi\sin\theta e^{i\phi} \\
0
\end{pmatrix}, 
~\bs{\psi}_{(3,4)}^{[0,0]} =\frac{1}{2}
\begin{pmatrix}
i\sin\xi\sin\chi\sin\theta e^{-i\phi} \\
\sin\xi (\cos\chi +i\sin\chi\cos\theta) \\
\sin\xi (\cos\chi-i\sin\chi\cos\theta) \\
-i\sin\xi \sin\chi\sin\theta e^{i\phi}
\end{pmatrix},
~\bs{\psi}_{(4,4)}^{[0,0]} =\frac{1}{\sqrt{2}}
\begin{pmatrix}
0 \\
i\sin\xi\sin\chi\sin\theta e^{-i\phi} \\
0 \\
\sin\xi (\cos\chi-i\sin\chi\cos\theta)
\end{pmatrix}.  \label{expeigenlowsii+i-1}
\end{align}
\normalsize
Equation 
(\ref{expeigenlowsii+i-1}) 
is realized as  a symmetric combination of the direct products of the monopole harmonics (\ref{uppertwocomp}) and the anti-monopole harmonics (\ref{lowertwocomp}):
\be
\bs{\psi}^{[0,0]}_{(\alpha,\beta)} =(\frac{1}{\sqrt{2}})^{\delta_{\alpha\beta}} ~(\bs{\psi}^{[0,\frac{1}{2}]}_{\alpha}\otimes \bs{\psi}^{[0,-\frac{1}{2}]}_{\beta} +  \bs{\psi}^{[0,\frac{1}{2}]}_{\beta}\otimes \bs{\psi}^{[0,-\frac{1}{2}]}_{\alpha}).  ~~~~(\alpha, \beta =1,2,3,4) \label{trssimplest}
\ee
With the $SO(5)$ charge conjugation matrix 
\be
{C} =\begin{pmatrix}
0 & 1 & 0 & 0 \\
-1 & 0 & 0 & 0 \\
0 & 0 & 0 & 1 \\
0 & 0 & -1 & 0 
\end{pmatrix}, 
\ee
we 
 see that (\ref{trssimplest}) is equivalent to $({C}\Sigma_{ab}^{(1,0)_5})_{\alpha\beta}\bs{\psi}^{[\frac{1}{2}, 0]}_{\alpha}\otimes \bs{\psi}^{[0, \frac{1}{2}]}_{\beta}$. 
  In (\ref{trssimplest}),  the monopole and  anti-monopole harmonics  equivalently  contribute to  the non-chiral monopole harmonics.  In the group theory point of view, Eq.(\ref{trssimplest}) corresponds to the symmetric $(2,0)_5$ representation made of two $(1,0)_5$ representations. Since the monopole harmonics and anti-monopole harmonics, respectively, have the $SU(2)$ gauge symmetry,  their tensor products (\ref{trssimplest}) enjoy the $SU(2)\otimes SU(2) \simeq SO(4)$ gauge symmetry. 
In general, the LLL non-chiral monopole harmonics in the $n=0$-sector of the $SO(4)$ monopole background  $(\frac{I}{4}, \frac{I}{4})_4$ $(I: \text{even  integers})$ can be obtained as the symmetric representation of the tensor product of two  LLL monopole harmonics of the $SU(2)$ monopole background $(\frac{I}{4}, 0)_4$ and the anti-monopole background $(0, \frac{I}{4})_4$:
\be
(\frac{I}{2}, 0)_5 ~\otimes ~(\frac{I}{2}, 0)_5 ~~\longrightarrow ~~(I, 0)_5. 
\ee





\begin{thebibliography}{99}



\bibitem{Yang-1978-1}
Chen Ning Yang, {\it``Generalization of Dirac's monopole to SU2 gauge fields''}, J. Math. Phys. 19 (1978) 320. 
\bibitem{Yang-1978-2}
Chen Ning Yang, {\it``SU2 monopole harmonics''}, J. Math. Phys. 19 (1978) 2622. 
\bibitem{Dirac-1931}
P.A.M. Dirac,      
{\it``Quantized singularities in the electromagnetic field''}, 
Proc. Royal Soc. London, A133 (1931)  60-72.
\bibitem{Zhang-Hu-2001}
S.C. Zhang and J.P. Hu,
\textit{``A four dimensional generalization of the quantum Hall effect''},
 Science 294 (2001) 823; cond-mat/0110572.
\bibitem{Hasebe-2020-1}
 Kazuki Hasebe,  
{\it ``$SO(5)$ Landau models and nested matrix geometry"}, 
 Nucl.Phys. B 956 (2020) 115012; arXiv:2002.05010. 
\bibitem{Hasebe-2014-1}
 Kazuki Hasebe,  
{\it ``Higher Dimensional Quantum Hall Effect as A-Class Topological Insulator"}, 
Nucl.Phys. B 886 (2014) 952-1002;  arXiv:1403.5066. 
\bibitem{Hasebe-2017}
 Kazuki Hasebe,  
{\it ``Higher (Odd) Dimensional Quantum Hall Effect and Extended Dimensional Hierarchy"},
Nucl.Phys. B 920 (2017) 475-520; arXiv:1612.05853.
\bibitem{WuZee1988} 
Y.S. Wu, A. Zee, 
{\it``Membranes, higher Hopf maps, and phase interactions''}, Phys. Lett. B {207} (1988) 39.  
\bibitem{TzeNam1989} 
C-H Tze, S. Nam, 
{\it``Topological phase entanglements of membrane solitons in division algebra sigma models with Hopf term''}, Annals  of  Phys. {193} (1989) 419.  
\bibitem{Hasebe-2020-3} 
Kazuki  Hasebe, 
{\it``A Unified Construction of Skyrme-type Non-linear sigma Models via The Higher Dimensional Landau Models''}, Nucl.Phys. B 961 (2020) 115250;  arXiv:2006.06152.  
\bibitem{Karabali-Nair-2002} 
Dimitra Karabali, V.P. Nair,  
{\it ``Quantum Hall Effect in Higher Dimensions''}, 
 Nucl.Phys. B641 (2002) 533-546;   
 hep-th/0203264. 
\bibitem{Bernevig-Hu-Toumbas-Zhang-2003}
 B. A. Bernevig, J. P. Hu, N. Toumbas, S. C. Zhang,
\textit{``The eight dimensional quantum Hall effect and the octonions''}, Phys.Rev.Lett. 91 (2003) 236803; 
cond-mat/0306045.
\bibitem{Hasebe-Kimura-2003}
K. Hasebe and Y. Kimura, 
{\it ``Dimensional Hierarchy in Quantum Hall Effects on Fuzzy Spheres''}, 
 Phys.Lett. B 602 (2004) 255; 
hep-th/0310274. 
\bibitem{Nair-Daemi-2004}
V.P. Nair, S. Randjbar-Daemi,    
{\it``Quantum Hall effect on $S^3$, edge states and fuzzy $S^3/{\bf Z}_2$''}, 
Nucl.Phys. B679 (2004) 447-463; hep-th/0309212.
\bibitem{Jellal-2005}
 A. Jellal,
{\it``Quantum Hall Effect on Higher Dimensional Spaces''}, Nucl.Phys. B725 (2005) 554-576;  
hep-th/0505095.
\bibitem{Hasebe-2010-2}
Kazuki Hasebe,   
{\it``Split-Quaternionic Hopf Map, Quantum Hall Effect, and Twistor Theory''}, 
Phys.Rev.D81 (2010) 041702; arXiv:0902.2523. 
\bibitem{Balli-Behtash-Kurkcuoglu-Unal-2014}
F. Balli, A. Behtash, S. Kurkcuoglu, G. Unal,     {\it``Quantum Hall Effect on the Grassmannians Gr2(CN)''}, 
	Phys. Rev. D 89 (2014) 105031;       arXiv:1403.3823. 
\bibitem{Hasebe-2014-2}
Kazuki Hasebe,  
{\it ``Chiral topological insulator on Nambu 3-algebraic geometry"},
 Nucl.Phys. B 886 (2014) 681-690;  arXiv:1403.7816. 
\bibitem{Karabali-Nair-2016} 
Dimitra Karabali, V.P. Nair,  
{\it ``The Geometry of Quantum Hall Effect: An Effective Action for all Dimensions''}, 
Phys. Rev. D 94 (2016) 024022;   arXiv:1604.00722. 
\bibitem{Coskun-Kurkcuoglu-Toga-2017}
U.H. Coskun, S. Kurkcuoglu, G.C.Toga,  
{\it ``Quantum Hall Effect on Odd Spheres"},
Phys. Rev. D 95 (2017) 065021; arXiv:1612.03855.
\bibitem{Heckman-Tizzano-2018}
Jonathan J. Heckman, Luigi Tizzano,     {\it``6D fractional quantum Hall effect''}, 
JHEP 05 (2018) 120-179;       arXiv:1708.02250. 
\bibitem{Hasebe-2004}
K. Hasebe,
{\it ``Supersymmetric Quantum Hall Effect on Fuzzy Supersphere''},
Phys.Rev.Lett. 94 (2005) 206802; hep-th/0411137. 
\bibitem{Hasebe-2008}
 Kazuki Hasebe
{\it ``Hyperbolic Supersymmetric Quantum Hall Effect''},
Phys.Rev.D78 (2008) 125024; arXiv:0809.4885.  
\bibitem{Sugawa-et-al-2018} 
S. Sugawa, F. Salces-Carcoba, A. R. Perry, Y. Yue, I. B. Spielman,   
{\it``Second Chern number of a quantum-simulated non-Abelian
Yang monopole''}, Science 360 (2018) 1429-1434. 
\bibitem{Ma-Bi-et-al-2021} 
Sh. Ma, Y. Bi, Q. Guo, B. Yang, O. You, J. Feng, H.-B. Sun, Sh. Zhang,    
{\it``Linked Weyl surfaces and Weyl arcs in photonic metamaterials''}, Science 373 (2021) 572-576. 
\bibitem{Price-et-al-2015}
H.M. Price, O. Zilberberg, T. Ozawa, I. Carusotto, N. Goldman,   
{\it``Four-Dimensional Quantum Hall Effect with Ultracold Atoms''}, 
Phys. Rev. Lett. 115 (2015) 195303. 
\bibitem{Price-et-al-2016}
H.M. Price, O. Zilberberg, T. Ozawa, I. Carusotto, N. Goldman, 
{\it``Measurement of Chern numbers through center-of-mass responses''}, 
Phys. Rev. B 93 (2016) 245113. 
\bibitem{Ozawa-Price-et-al-2016}
 T. Ozawa, H.M. Price,N. Goldman, O. Zilberberg, I. Carusotto,      
{\it``Synthetic dimensions in integrated photonics:
From optical isolation to four-dimensional quantum Hall physics''}, 
Phys. Rev. A 93 (2016) 043827. 
\bibitem{Wang-Price-Zhang-Chong-2020}
You Wang, Hannah M. Price, Baile Zhang, Y. D. Chong, {\it``Circuit implementation of a four-dimensional topological insulator''}, Nature Communications, 11 (2020) 2356; arXiv:2001.07427.  
\bibitem{Chen-Zhu-Tan-Wang-Ma-2021} 
Ze-Guo Chen, Weiwei Zhu, Yang Tan, Licheng Wang, and Guancong Ma, {\it``Acoustic Realization of a Four-Dimensional Higher-Order Chern Insulator and Boundary-Modes Engineering''}, Phys. Rev. X 11 (2021) 011016; arXiv:1912.10267.   
\bibitem{Lohse-et-al-2018} 
M. Lohse, Ch. Schweizer, H. M. Price, O. Zilberberg, I. Bloch,   
{\it``Exploring 4D quantum Hall physics with a 2D
topological charge pump''}, Nature 553 (2018) 55. 
\bibitem{Zilberberg-et-al-2018} 
O. Zilberberg, Sh. Huang, J. Guglielmon, M. Wang, K.P. Chen, Y.E. Kraus, M.C. Rechtsman,    
{\it``Photonic topological boundary pumping as a probe of 4D quantum Hall physics''}, Nature 553  (2018) 59. 
\bibitem{Lian-Zhang-2016}
Biao Lian, Shou-Cheng Zhang,   
{\it``A Five Dimensional Generalization of the Topological Weyl Semimetal''}, 
Phys Rev B 94 (2016) 041105;   arXiv:1604.07459. 
\bibitem{Lian-Zhang-2017}
Biao Lian, Shou-Cheng Zhang,   
{\it``Weyl semimetal and topological phase transition in five dimensions''}, 
Phys Rev B 95 (2017) 235106;   arXiv:1702.07982. 
\bibitem{Hashimoto-Kimura-Wu-2017}
Koji Hashimoto, Taro Kimura, Xi Wu,    
{\it``Edge-of-edge states''}, 
Phys. Rev. B 95 (2017) 165443;    arXiv:1702.00624. 
\bibitem{Hashimoto-Matsuo-2020}
Koji Hashimoto, Yoshinori Matsuo,   
{\it``Universal higher-order topology from a five-dimensional Weyl semimetal: Edge topology, edge Hamiltonian, and a nested Wilson loop''}, 
Phys Rev B 101 (2020) 245138;    arXiv:2002.12596. 
\bibitem{Kane-Mele-2005}
C.L. Kane, E.J. Mele, 
{\it``Z$_2$ Topological Order and the Quantum Spin Hall Effect''}, 
Phys. Rev. Lett. 95 (2005) 146802, cond-mat/0506581. 
\bibitem{Bernevig-Zhang-2005}
B. Andrei Bernevig, Shou-Cheng Zhang, 
{\it``Quantum Spin Hall Effect''}, 
Phys. Rev. Lett. 96 (2006) 106802, cond-mat/0504147. 
\bibitem{Sheng-Weng-Sheng-Haldane-2006}
D.N. Sheng, Z.Y. Weng, L. Sheng, F.D.M. Haldane, 
{\it``Quantum Spin-Hall Effect and Topologically Invariant Chern Numbers''}, 
Phys. Rev. Lett. 97 (2006) 036808, cond-mat/0603054. 
\bibitem{Li-Wu-2013}
Yi Li, Congjun Wu, 
{\it``High-Dimensional Topological Insulators with Quaternionic Analytic Landau Levels''}, 
Phys. Rev. Lett. 110 (2013) 216802; arXiv:1103.5422. 
\bibitem{Li-Zhang-Wu-2013}
Yi Li, Shou-Cheng Zhang, Congjun Wu, 
{\it``Topological insulators with SU(2) Landau levels''},  
 	Phys. Rev. Lett. 111 (2013) 186803;  	arXiv:1208.1562. 
\bibitem{Ryu-Takayanagi-2010-1}
Shinsei Ryu, Tadashi Takayanagi, {\it``Topological Insulators and Superconductors from D-branes''}, 
Phys.Lett.B 693 (2010) 175-179;      
 arXiv:1001.0763. 
\bibitem{Ryu-Takayanagi-2010-2}
Shinsei Ryu, Tadashi Takayanagi,  {\it``Topological Insulators and Superconductors from String Theory''}, 
Phys.Rev.D 82 (2010) 086014;      
 arXiv:1007.4234. 
\bibitem{Hasebe-2010}
 Kazuki Hasebe,  
{\it ``Hopf Maps, Lowest Landau Level, and Fuzzy Spheres"},
SIGMA 6 (2010) 071; arXiv:1009.1192.
\bibitem{Karabali-Nair-review} 
Dimitra Karabali, V.P. Nair,  
{\it ``Quantum Hall effect in higher dimensions, matrix models and fuzzy geometry''}, 
 Jour. Phys. A: Math. Gen. 39 (2006) 12735-12763;   hep-th/0606161. 
\bibitem{Coleman-Wess-Zumino-1969}
S. Coleman, J. Wess, B. Zumino,   
{\it ``Structure of Phenomenological Lagrangians. I''}, 
 Phys. Rev. 177 (1969) 2239-2246.  
\bibitem{CallanJr-Coleman-Wess-Zumino-1969}
C. G. Callan, Jr., S. Coleman, J. Wess, B. Zumino,   
{\it ``Structure of Phenomenological Lagrangians. II''}, 
 Phys. Rev. 177 (1969) 2247-2250.  
\bibitem{Salam-Strathdee-1982}
Abdus Salam, J. Strathdee,  
{\it ``On Kaluza-Klein Theory''}, 
 Ann. Phys. 141 (1982) 316-352.  
\bibitem{Nair-book}  
V.P. Nair,  
{\it ``Quantum Field Theory: A Modern Perspective''}, 
 Springer (2005). 
\bibitem{Hasebe-2015}
 Kazuki Hasebe,  
{\it ``Relativistic Landau Models and Generation of Fuzzy Spheres"},
 Int.J.Mod.Phys.A 31 (2016) 1650117;  arXiv:1511.04681. 
\bibitem{Hasebe-2018} 
Kazuki  Hasebe, 
{\it``$SO(4)$ Landau Models and Matrix Geometry''}, Nucl.Phys. B 934 (2018) 149-211;  arXiv:1712.07767. 
\bibitem{Wu-Yang-1976}
T.T. Wu, C.N. Yang,      
{\it``Dirac Monopoles without Strings: Monopole Harmonics''}, 
Nucl.Phys. B107 (1976) 365-380.
\bibitem{Shnir-book}
Yakov M. Shnir,      
{\it``Magnetic Monopoles''},  
Springer (2005). 
\bibitem{Haldane-1983} 
F.D.M. Haldane,
{\it``Fractional quantization of the Hall effect: a hierarchy of incompressible
quantum fluid  states"}, 
 Phys. Rev. Lett. 51 (1983) 605-608. 
\bibitem{BPST-1975}
A. A. Belavin, A. M. Polyakov, A. S. Schwartz, Yu. S. Tyupkin,      
{\it``Pseudoparticle solutions of the Yang-Mills equations''},  
Phys.Lett.B 59 (1975) 85-87. 
\bibitem{Jackiw-Rebbi-1976}
R. Jackiw and C. Rebbi,
{\it``Conformal properties of a Yang-Mills pseudoparticle''}, 
Phys. Rev.D  {\bf 14}   (1976) 517.
\bibitem{Eguchi-Freund-1976}
Tohru Eguchi, Peter G. O. Freund,   
{\it``Quantum Gravity and World Topology''}, 
Phys. Rev. Lett. 37  (1976) 1251. 
\bibitem{Ishiki-Matsumoto-Muraki-2018}
G. Ishiki, T. Matsumoto, H. Muraki,  
{\it``Information metric, Berry connection, and Berezin-Toeplitz quantization for matrix geometry''},  
 Phys. Rev. D 98 (2018) 026002;   arXiv:1804.00900. 
\bibitem{Girvin-Jach-1984}
S.M. Girvin, Terrence Jach,  
{\it``Formalism for the quantum Hall effect: Hilbert space of analytic functions''},  
 Phys. Rev. B 29 (1984) 5617-5625.  
\bibitem{Girvin-MacDonald-Platzman-1986}
S.M. Girvin, A.H. MacDonald, P.M. Platzman,     
{\it``Magneto-roton theory of collective excitations in the fractional quantum Hall effect''},  
 Phys. Rev. B 33 (1986) 2481-2494.  
\bibitem{Ezawa-Tsitsishvili-Hasebe-2003}
Z.F. Ezawa, G. Tsitshishvili, K. Hasebe,   
{\it``Noncommutative geometry, extended W$_\infty$ algebra, and Grassmannian solitons in multicomponent quantum Hall systems''},  
 Phys. Rev. B 67 (2003) 125314.  
\bibitem{Dolan-2003} 
Brian P. Dolan, 
{\it``The Spectrum of the Dirac Operator on Coset Spaces with Homogeneous Gauge Fields''}, JHEP 0305 (2003) 018;  hep-th/0304037. 




\end{thebibliography}
\end{document}